\documentclass[a4paper,11pt]{article}
\usepackage{jheppub}

\pdfoutput=1
\usepackage{amsmath}
\usepackage{amsfonts}
\usepackage{comment}
\usepackage{amssymb}
\usepackage{mathrsfs}
\usepackage{graphicx}
\usepackage{color}
\usepackage{multirow}
\usepackage{bm}
\usepackage{tabularx}
\usepackage{subcaption}
\usepackage{blindtext}
\usepackage{wasysym}
\usepackage{hyperref}
\usepackage{float}
\hypersetup{colorlinks=true,allcolors=blue}
\usepackage{orcidlink}
\usepackage[normalem]{ulem}
\usepackage{lineno}
\usepackage[T1]{fontenc}
\usepackage{soul}
\usepackage[normalem]{ulem}
\usepackage{multirow}
\usepackage{array}
\newcolumntype{L}{>{\centering\arraybackslash}m{1.99cm}}

\newcommand{\bi}{\begin{itemize}}
\newcommand{\ei}{\end{itemize}}

\newcommand{\gsol}{$\Gamma_{21}$~}
\newcommand{\gatm}{$\Gamma_{32}$~}

\newcommand{\ess}{ESSnuSB~}
\newcommand{\dcp}{$\delta_{\rm CP}$}

\begin{document}
 
\title{Decoherence in Neutrino Oscillation at the ESSnuSB Experiment
\\ \vspace{4mm}
{\small (ESSnuSB Collaboration)}}

\newcommand{\authorlist}{

\author[1]{J.~Aguilar,}               
\author[2]{M.~Anastasopoulos,}        
\author[3]{E.~Baussan,}               
\author[2]{A.K.~Bhattacharyya,}       
\author[2]{A.~Bignami,}               
\author[4,5]{M.~Blennow,}             
\author[6]{M.~Bogomilov,}             
\author[2]{B.~Bolling,}               
\author[3]{E.~Bouquerel,}             
\author[7]{F.~Bramati,}               
\author[7]{A.~Branca,} 
\author[7]{G.~Brunetti,}                
\author[1]{I.~Bustinduy,}             
\author[8]{C.J.~Carlile,}             
\author[8]{J.~Cederkall,}             
\author[9]{T.~W.~Choi,}               
\author[4,5]{S.~Choubey,}             
\author[8]{P.~Christiansen,}          
\author[2,10]{M.~Collins,}            
\author[7]{E.~Cristaldo Morales,}     
\author[11]{P.~Cupia{\l},}               
\author[2]{H.~Danared,}
\author[9]{D.~Dancila,}               
\author[3]{J.~P.~A.~M.~de~Andr\'{e},} 
\author[3]{M.~Dracos,}                
\author[12]{I.~Efthymiopoulos,}       
\author[9]{T.~Ekel\"{o}f,}            
\author[2]{M.~Eshraqi,}
\author[13]{G.~Fanourakis,}
\author[14]{A.~Farricker,}
\author[15]{E.~Fasoula,}
\author[16]{T.~Fukuda,}
\author[2]{N.~Gazis,}
\author[13]{Th.~Geralis,}
\author[17,*]{M.~Ghosh\orcidlink{0000-0003-3540-6548},\note[*]{Corresponding authors}}
\author[18,*]{A.~Giarnetti\orcidlink{0000-0001-8487-8045},}
\author[19,20]{G.~Gokbulut}
\author[21,22,*]{A.~Gupta\orcidlink{0000-0002-7247-2424},}
\author[23]{C.~Hagner,}
\author[17]{L.~Halić,}
\author[23]{V.~T.~Hariharan,}
\author[20]{M.~Hooft,}
\author[8]{K.~E.~Iversen,}
\author[20]{N.~Jachowicz,}
\author[2]{M.~Jenssen,}
\author[2]{R.~Johansson,}
\author[15]{E.~Kasimi,}
\author[19]{A.~Kayis Topaksu,}
\author[2]{B.~Kildetof,}
\author[17]{B.~Kliček,}
\author[15]{K.~Kordas,}
\author[24]{A.~Leisos,}
\author[2,8]{M.~Lindroos,}
\author[25]{A.~Longhin,}
\author[2]{C.~Maiano,}
\author[21,22]{D.~Majumdar,}
\author[7]{S.~Marangoni,}
\author[2]{C.~Marrelli,}
\author[2]{C.~Martins,}
\author[18,*]{D.~Meloni\orcidlink{0000-0001-7680-6957},}
\author[26]{M.~Mezzetto,}
\author[2]{N.~Milas,}
\author[1]{J.~Muñoz,}
\author[20]{K.~Niewczas,}
\author[19]{M.~Oglakci,}
\author[4,5]{T.~Ohlsson,}
\author[9]{M.~Olveg\r{a}rd,}
\author[25]{M.~Pari,}
\author[2]{D.~Patrzalek,}
\author[6]{G.~Petkov,}
\author[15]{Ch.~Petridou,}
\author[3]{P.~Poussot,}
\author[13]{A.~Psallidas,}
\author[26]{F.~Pupilli,}
\author[27]{D.~Saiang,}
\author[15]{D.~Sampsonidis,}
\author[3]{C.~Schwab,}
\author[1]{F.~Sordo,}
\author[2]{A.~Sosa,}
\author[13]{G.~Stavropoulos,}
\author[17]{M.~Stipčević,}
\author[2]{R.~Tarkeshian,}
\author[7]{F.~Terranova,}
\author[23]{T.~Tolba,}
\author[2]{E.~Trachanas,}
\author[6]{R.~Tsenov,}
\author[24]{A.~Tsirigotis,}
\author[15]{S.~E.~Tzamarias,}
\author[6]{G.~Vankova-Kirilova,}
\author[28]{N.~Vassilopoulos,}
\author[4,5]{S.~Vihonen,}
\author[3]{J.~Wurtz,}
\author[3]{V.~Zeter,}
\author[13]{O.~Zormpa,}
\author[9]{and Y.~Zou}

\affiliation[1]{Consorcio ESS-bilbao, Parque Científico y Tecnológico de Bizkaia, Laida Bidea, Edificio 207-B, 48160 Derio, Bizkaia}
\affiliation[2]{European Spallation Source, Box 176, SE-221 00 Lund, Sweden}
\affiliation[3]{IPHC, Universit\'{e} de Strasbourg, CNRS/IN2P3, F-67037 Strasbourg, France}
\affiliation[4]{Department of Physics, School of Engineering Sciences, KTH Royal Institute of Technology, Roslagstullsbacken 21, 106 91 Stockholm, Sweden}
\affiliation[5]{The Oskar Klein Centre, AlbaNova University Center, Roslagstullsbacken 21, 106 91 Stockholm, Sweden}
\affiliation[6]{Sofia University St. Kliment Ohridski, Faculty of Physics, 1164 Sofia, Bulgaria}
\affiliation[7]{University of Milano-Bicocca and INFN Sez. di Milano-Bicocca, 20126 Milano, Italy}
\affiliation[8]{Department of Physics, Lund University, P.O Box 118, 221 00 Lund, Sweden}
\affiliation[9]{Department of Physics and Astronomy, FREIA Division, Uppsala University, P.O. Box 516, 751 20 Uppsala, Sweden}
\affiliation[10]{Faculty of Engineering, Lund University, P.O Box 118, 221 00 Lund, Sweden}
\affiliation[11]{AGH University of Krakow, al. A. Mickiewicza 30, 30-059 Krakow, Poland}
\affiliation[12]{CERN, 1211 Geneva 23, Switzerland}
\affiliation[13]{Institute of Nuclear and Particle Physics, NCSR Demokritos, Neapoleos 27, 15341 Agia Paraskevi, Greece}
\affiliation[14]{Cockroft Institute (A36), Liverpool University, Warrington WA4 4AD, UK}
\affiliation[15]{Department of Physics, Aristotle University of Thessaloniki, Thessaloniki, Greece}
\affiliation[16]{Institute for Advanced Research, Nagoya University, Nagoya 464–8601, Japan}
\affiliation[17]{Center of Excellence for Advanced Materials and Sensing Devices, Ru{\dj}er Bo\v{s}kovi\'c Institute, 10000 Zagreb, Croatia}
\affiliation[18]{Dipartimento di Matematica e Fisica, Università di Roma Tre Via della Vasca Navale 84, 00146 Rome, Italy}
\affiliation[19]{University of Cukurova, Faculty of Science and Letters, Department of Physics, 01330 Adana, Turkey}
\affiliation[20]{Department of Physics and Astronomy, Ghent University, Proeftuinstraat 86, B-9000 Ghent, Belgium}
\affiliation[21]{Theory Division, Saha Institute of Nuclear Physics, 1/AF, Bidhannagar, Kolkata 700064, India}
\affiliation[22]{Homi  Bhabha  National  Institute,  Anushakti  Nagar,  Mumbai  400094,  India}
\affiliation[23]{Institute for Experimental Physics, Hamburg University, 22761 Hamburg, Germany}
\affiliation[24]{Physics Laboratory, School of Science and Technology, Hellenic Open University, 26335, Patras, Greece}
\affiliation[25]{Department of Physics and Astronomy "G. Galilei", University of Padova and INFN Sezione di Padova, Italy}
\affiliation[26]{INFN Sez. di Padova, Padova, Italy}
\affiliation[27]{Department of Civil, Environmental and Natural Resources Engineering $Lule\aa~University~of~Technology$, SE-971 87 Lulea, Sweden}
\affiliation[28]{Institute of High Energy Physics (IHEP) Dongguan Campus, Chinese Academy of Sciences (CAS), 1 Zhongziyuan Road, Dongguan, Guangdong, 523803, China}

\emailAdd{mghosh@irb.hr}
\emailAdd{alessio.giarnetti@uniroma3.it}
\emailAdd{aman.gupta@saha.ac.in}
\emailAdd{davide.meloni@uniroma3.it}




}

\authorlist

\abstract{Neutrino oscillation experiments provide a unique window in exploring several new physics scenarios beyond the standard three flavour. One such scenario is quantum decoherence in neutrino oscillation which tends to destroy the interference pattern of neutrinos reaching the far detector from the source. In this work, we study the decoherence in neutrino oscillation in the context of the ESSnuSB experiment. We consider the energy-independent decoherence parameter and derive the analytical expressions for P$_{\mu e}$ and P$_{\mu \mu}$ probabilities in vacuum. We have computed the capability of ESSnuSB to put bounds on the decoherence parameters namely,  $\Gamma_{21}$ and $\Gamma_{32}$ and found that the constraints on $\Gamma_{21}$ are competitive compared to the DUNE bounds and better than the most stringent LBL ones from MINOS/MINOS+. We have also investigated the impact of decoherence on the ESSnuSB measurement of the Dirac CP phase $\delta_{\rm CP}$ and concluded that it remains robust in the presence of new physics.  
}


\maketitle

\section{Introduction}
The discovery of atmospheric \cite{Super-Kamiokande:1998kpq, Kajita:2016cak} and solar \cite{SNO:2002tuh, McDonald:2016ixn} neutrino oscillations has firmly confirmed the theory of neutrino oscillation, first proposed by Pontecorvo  \cite{Gribov:1968kq, Bilenky:1978nj}. It is considered as the quantum mechanical interference phenomenon governed by the three mixing angles: $\theta_{23}$, $\theta_{13}$, $\theta_{12}$, two independent mass-squared splittings: $\Delta m^2_{21}$, $\Delta m^2_{31}$ and the leptonic CP violating phase $\delta_{\rm CP}$. Out of six parameters, the measurement of $\delta_{\rm CP}$ might help us to solve the problem of matter-antimatter asymmetry of the Universe \cite{Gavela:1993ts} that we observe today.  So far, there is no conclusive evidence of the CP symmetry violation (CPV) in neutrino oscillation although hints for maximal violation are emerging \cite{T2K:2023smv}.  One of the primary goals of current and forthcoming oscillation experiments \cite{t2k:2019bcf, T2K:2017hed, Acero_2019, DUNE_1} is to measure the possible value of $\delta_{\rm CP}$ with utmost precision. The European Spallation Source (ESS) neutrino Super-Beam \ess \cite{ESSnuSB:2021azq} is a next-to-next generation accelerator-based long-baseline neutrino oscillation experiments in Sweden which uses the second oscillation maximum in the appearance probability P$_{\mu e}$ in order to measure $\delta_{\rm CP}$. In this experiment, a high-intensity muon neutrino beam will be produced using a 5 MW proton beam with the upgraded ESS facility in Lund \cite{Abele:2022iml, Alekou:2022emd}. These neutrinos, then, will be detected by a water Cherenkov detector located at a far distance of 360 km at Zinkgruvan mine (see the conceptual design report (CDR) \cite{Alekou:2022emd} for more details). \\
In this work, we study the quantum decoherence effects in neutrino oscillation in the context of \ess experiment. According to the neutrino oscillation model, neutrino flavour states (also known as weak interaction states) are not the same as mass eigenstates (also known as propagation states), so that a flavour state can be seen as a linear superposition of different mass eigenstates. During their propagation, the latter evolve coherently with different frequencies, giving rise to the phenomenon of neutrino flavour conversion. The relevant point here is that different mass eigenstates maintain their relative phase differences as they propagate. However, there exist several mechanisms that lead to the destruction of such interference patterns, and coherence in different neutrino mass eigenstates may get lost. One such a mechanism is the wave packet separation where the coherence is lost among different neutrino mass states owing to their different group velocities after traveling over long distances. This can be described by the usual quantum mechanical framework and has been studied in detail in Refs.  \cite{Giunti:1997wq,Ohlsson:2000mj,deGouvea:2020hfl,Blasone:2021cau,Luciano:2021rgt,Akhmedov:2022bjs,Jones:2022hme, Marzec:2022mcz,Ciuffoli:2022uzf,Guzzo:2014jbp}. \\
There is another general treatment of decoherence effects which considers an open quantum system framework~\cite{Breuer:2002pc} and uses density matrix formalism. In this method, one describes the neutrinos as a subsystem interacting with the environment, causing the dissipative effects in the neutrino oscillation phenomenon which have been discussed in a variety of oscillation experiments ~\cite{Benatti:2000ph,Lisi:2000zt, Adler:2000vfa,Benatti:2001fa, Gago:2000qc, Hooper:2004xr,Anchordoqui:2005gj,Fogli:2007tx,Farzan:2008zv, Jones:2014sfa,  BalieiroGomes:2016ykp, Coelho:2017zes, Coelho:2017byq, Carpio:2017nui,   Coloma:2018idr, Ohlsson:2020gxx,Stuttard:2020qfv,Stuttard:2021uyw, Carpio:2018gum, Carrasco:2018sca, Buoninfante:2020iyr, Gomes:2020muc, Capolupo:2020enx}. The effect of environmentally induced decoherence is to mainly introduce damping terms in the oscillation probabilities which, in general, can be energy-dependent. In earlier works, the energy-dependent constraints on decoherence parameters have been obtained in various experiments that include, among others,  IceCube~\cite{Coloma:2018idr}, Super-Kamiokande~\cite{Lisi:2000zt, Fogli:2007tx}, KamLAND~\cite{deOliveira:2013dia}, MINOS~\cite{BalieiroGomes:2016ykp}, T2K~\cite{Gomes:2020muc}, NOvA~\cite{Coelho:2017zes}. On the other hand, bounds on the energy-independent decoherence parameters have been obtained from T2K and MINOS~\cite{Gomes:2020muc}, from the future DUNE experiment \cite{BalieiroGomes:2018gtd, Carpio:2018gum} and from solar \cite{ deHolanda:2019tuf} and reactor \cite{Marzec:2022mcz,deGouvea:2020hfl} neutrinos. The impact of decoherence on the precision measurements at DUNE and T2HK has been examined in \cite{Barenboim:2024wdn}. Quantum decoherence may also be induced by stochastic metric fluctuations affecting the neutrino oscillations as shown in Ref.~\cite{DEsposito:2023psn}. In the recent work \cite{Domi:2024ypm}, the authors delve into the exploration of gravitationally induced decoherence, providing a comprehensive analysis that includes a comparison with various phenomenological models. It is worthwhile to note that, in addition to the dissipative characteristics of environmentally induced decoherence, the open quantum system framework also enables the exchange of energy between the neutrino sub-system and the environmental field. In this scenario, one may observe both relaxation and quantum decoherence effects \cite{Guzzo:2014jbp}, a phenomenon recently investigated in Refs. \cite{Gomes:2020muc,Barenboim:2024zfi}. However, our primary emphasis here is on the dissipative nature of decoherence as the main observable. \\

In the present work, to the best of our knowledge, we study for the first time the effects of neutrino decoherence in the standard three flavour oscillation picture in the context of the \ess experiment and present the bounds on {\it energy-independent} decoherence parameters that such a facility can provide, illustrating the main differences with respect to similar bounds achievable in DUNE \cite{BalieiroGomes:2018gtd, Carpio:2018gum}. We also investigate the impact of decoherence on the $\delta_{\rm CP}$ measurement of ESSnuSB. Our numerical results are easily understood by means of simple analytical expressions for oscillation probabilities in the presence of decoherence. \\

This paper is structured in the following manner. In the next section, we provide a brief overview of the decoherence formalism in neutrino oscillation,  considering the open quantum-system approach and derive the related analytical formulae. The description of the \ess experiment and other simulation details are given in section 3. In section 4, we compute the transition probabilities and generate the event plots in the presence of decoherence for \ess. Finally, the sensitivity of the \ess experiment to constrain the decoherence parameters and their impact on the measurement of $\delta_{\rm CP}$ are illustrated in section 5.

\section{Formalism}
\label{sec:formal}
In addition to the usual Schrodinger wave mechanics method, the neutrino oscillation formalism can also be understood using the density matrix approach. Considering neutrinos as an open quantum system interacting with the surroundings, their evolution equation is given by the Lindblad Master equation \cite{Lindblad:1975ef,Gorini:1975nb,Oliveira:2010zzd}  

\begin{equation}
	\frac{\partial\rho(t)}{\partial t}=-i[H,\rho(t)] + \mathcal{D}[\rho(t)]\,,
    \label{eq:lindblad}
\end{equation}
where $\rho(t)$ is the density matrix corresponding to the neutrino states and $H$ is the neutrino (subsystem) Hamiltonian which, in the presence of ambient matter, can be written in  the flavour basis as

\begin{align}
    H & = \frac{1}{2E}\left[U \begin{pmatrix} 
0 & 0 & 0 \\
0 & \Delta m^2_{21} & 0 \\
0 & 0 & \Delta m^2_{31}
\end{pmatrix} U^\dagger + 2EV_{CC}
\begin{pmatrix} 
1 & 0& 0 \\
0 & 0& 0 \\
0 & 0 & 0
\end{pmatrix}\right]
\label{eq:Hamiltonian}.
\end{align}
\\
In the previous expression, $U$ is the standard Pontecorvo-Maki-Nakagawa-Sakata (PMNS) mixing matrix in vacuum, and $V_{CC} = \pm \sqrt{2}G_F N_e$ is the matter potential term due to the CC interactions of neutrinos with matter. The neutrino energy is denoted by $E$, and $N_e$ is the electron number density.
The effect of decoherence is given by the dissipator $\mathcal{D}$ which can be expressed in terms of the Lindblad dissipative operators for $N$ dimension $\mathcal{L}$ as ~\cite{Gago:2002na,Benatti:2000ph,Lisi:2000zt}
\begin{equation}
    \label{eq:dissipative}
    \mathcal{D}[\rho(t)] = \frac{1}{2}\sum_{i=1}^{N^2-1} \left([\mathcal{L}_i,\rho(t) \mathcal{L}_i^{\dagger}]+[\mathcal{L}_i\rho(t), \mathcal{L}_i^{\dagger}] \right) \, .
\end{equation}
It should be noted that such Lindblad operators $\mathcal{L}$ and hence dissipator $\mathcal{D}$ can be expanded in terms of $SU(N)$ generators \cite{Buoninfante:2020iyr,Stuttard:2020qfv} involving a large number of free degrees of freedom. The number of (free) parameters is reduced by imposing several physical constraints on $\mathcal{D}$ such as unitarity, complete positivity, entropy increase and energy conservation (for a detailed discussion see ~\cite{Lisi:2000zt,Benatti:2000ph,Oliveira:2010zzd,BalieiroGomes:2018gtd, DeRomeri:2023dht}). Under such assumptions, 
in the present analysis, we have used the simplest form of the dissipative matrix $\mathcal{D}$ which contains the decoherence parameters affecting the neutrino oscillation probabilities. This matrix, in the three neutrino case can be expanded as $\mathcal{D}=D_{jk}\rho_k \lambda_j$, where $\lambda_j$ are the Gell-Mann matrices and $\rho_k$ are the elements of the neutrino density matrix. By imposing the above-mentioned physical conditions the dissipator takes the form \cite{Oliveira:2013nua,BalieiroGomes:2016ykp,BalieiroGomes:2018gtd, Oliveira:2016asf}

\begin{equation}
 \label{eq:Dmatrixdiag}
	D_{jk}= -\mathrm{diag}(\Gamma_{21},\Gamma_{21},0,\Gamma_{31}, \Gamma_{31},\Gamma_{32},\Gamma_{32},0)\,.
\end{equation}

The net effect of decoherence is to introduce terms similar to damping phenomena of the form $e^{-\Gamma_{ij} L}$ in the oscillation probability, where $L$ is the baseline for neutrino oscillation which is $360$ km for the \ess experiment. 

It should be noted here that for the sake of simplicity, we only consider energy-independent decoherence matrix elements \footnote{For the effects of energy-dependent decoherence elements we refer to \cite{Coloma:2018idr, DeRomeri:2023dht}. We expect that for $\Gamma_{ij}\propto E^{n}$, the ESSnuSB limits on decoherence parameters should improve if $n<0$ and worsen for $n>0$ due to the neutrino low energy spectrum. However, this is beyond the scope of the present work and it might be possible to investigate such scenarios in a future study.}; in such a scenario, only two $\Gamma_{ij}$ are independent because of the relation \cite{Oliveira:2016asf,BalieiroGomes:2018gtd}
\begin{align}
    \Gamma_{31} = \Gamma_{21} + \Gamma_{32} - 2\sqrt{\Gamma_{21}\Gamma_{32}}\,.
    \label{eq:gamma_relation}
\end{align}

In our analysis, we will present all our results in terms of the two independent parameters \gatm and \gsol. Note that between these two parameters, if we take \gatm to be non-zero and \gsol = 0, then it will imply \gatm $= \Gamma_{31}$ and if we take \gsol to be non-zero and \gatm = 0 then it will imply \gsol $= \Gamma_{31}$. Further, if we consider \gatm = \gsol, then it implies $\Gamma_{31} = 0$. These conditions will be crucial for interpreting the results of our analysis. 

The neutrino oscillation probability including the decoherence effect considered here is then given by \cite{Blennow:2005yk,BalieiroGomes:2018gtd}

\begin{align}
    P(\nu_\alpha\to\nu_\beta)=\delta_{\alpha\beta} &- 2\sum_{i>j} {\rm Re}\left[ \tilde{U}^*_{\alpha i}\tilde{U}_{\beta i}\tilde{U}_{\beta j}\tilde{U}^*_{\beta j} \right] ~\left[1-\cos \left(2\tilde{\Delta}_{ij} \right)~e^{-\Gamma_{ij}L}\right]\nonumber
    \\
    &+2\sum_{i>j}{\rm Im}\left[ \tilde{U}^*_{\alpha k}\tilde{U}_{\beta k}\tilde{U}_{\beta j}\tilde{U}^*_{\beta j} \right] ~\sin \left(2\tilde{\Delta}_{ij}\right)~e^{-\Gamma_{ij}L}\,,
    \label{eq:finalprob}
\end{align}
where $\tilde{U}$ is the modified PMNS matrix in matter and $\tilde{\Delta}_{ij} = \dfrac{\Delta\tilde{m}^2_{ij}L}{4E}$, with $\Delta\tilde{m}^2_{ij}$ being the mass squared differences in the presence of matter. Our choice of decoherence formalism is motivated by the fact that it is easy to understand which oscillation frequency is attenuated by the decoherence parameter $\Gamma_{ij}$. Switching on $\Gamma_{21}$ exclusively, for instance, we observe the suppression of the oscillation frequency associated with the solar mass-squared difference ($\Delta m^2_{21}$) while the oscillation frequency corresponding to the atmospheric mass-squared difference ($\Delta m^2_{31}$) remains unaffected. This model also facilitates a direct comparison between the other experimental bounds on $\Gamma_{ij}$ and those derived from our present work. We would like to emphasize that, while the matrix $D$ is conventionally defined in vacuum, the inclusion of matter effects necessitates a rotation of the decoherence matrix $D$ to the matter basis \cite{Carpio:2017nui}. Consequently, $D$ ceases to be a diagonal matrix, thereby modifying eq. \ref{eq:finalprob} accordingly. However, given that matter effects for the \ess are relatively small compared to atmospheric and astrophysical neutrinos, the approximate formula (eq. \ref{eq:finalprob}) remains valid and appropriate for our current analysis \cite{DeRomeri:2023dht}. Furthermore, the influence of non-diagonal elements is significant only at large neutrino energies, which is not pertinent to the scope of our work. For a more detailed discussion about the validity of eq. \ref{eq:finalprob} we refer to Appendix B of Ref.~\cite{BalieiroGomes:2018gtd}
. In the context of long-baseline (LBL) experiments, the most stringent current bounds have been obtained by the MINOS/MINOS+ data recent analysis from \cite{DeRomeri:2023dht}
\begin{eqnarray}
    \Gamma_{32}=\Gamma_{21}&<&9.4\times 10^{-24} \, \, \mathrm{GeV} \, \, \, \, \mathrm{[MINOS/MINOS+, (90\% \rm ~C.L.)],} \label{bounds:MINOS}
\end{eqnarray}
while DUNE is expected to reach with its standard neutrino flux \cite{BalieiroGomes:2018gtd}
\begin{eqnarray}
    \Gamma_{21}&<&1.2 \times 10^{-23} \, \, \mathrm{GeV} \, \, \, \, \mathrm{[DUNE, (90\% \rm ~C.L.)]} \\
    \Gamma_{32}&<&4.7 \times 10^{-24} \, \, \mathrm{GeV} \, \, \, \, \mathrm{[DUNE, (90\% \rm ~C.L.)]} \,.\label{bounds:dune}
\end{eqnarray}
It is worth mentioning that with a high energy flux, DUNE might improve the \gatm bound up to  $ \Gamma_{32}<7.7 \times 10^{-25}$ GeV ($90\% ~\rm C.L.$) \cite{BalieiroGomes:2018gtd}.

\subsection{Oscillation Probability in Vacuum}
\label{sec:prob:vacuum}
We present here the oscillation probabilities relevant for LBL experiments, namely the electron neutrino appearance and the muon neutrino disappearance. We write the probabilities as
\begin{equation}
    P_{\alpha\beta}=P_{\alpha\beta}^{SM}+P_{\alpha\beta}^{deco},
\end{equation}
where the first term depicts the standard oscillations while the second refers to the decoherence correction. We expand up to the second order in $\sin\theta_{13}$ and in $\alpha=\Delta m^2_{21}/\Delta m^2_{31}$ and up to the first order in the small quantities dependent on the decoherence parameters $\Gamma_{21}L$ and $\Gamma_{32}L$. In this analytic approach, we neglect the standard matter effects, since they are not relevant under ESSnuSB conditions and would make the oscillation probabilities less readable. In the appearance case, we get:
\begin{align}
\nonumber P_{\mu e}^{SM}&= 4 s_{13}^2s_{23}^2\sin^2{\Delta_{31}} + 2\alpha \Delta_{31}s_{13}\sin{\Delta_{31}}\sin{2\theta_{12}}\sin{2\theta_{23}}\cos({\delta_{\rm CP} + \Delta_{31}})
 +(\alpha\Delta_{31}c_{12}c_{23}s_{12})^2,\\ \nonumber \\ \nonumber
P_{\mu e}^{deco}&=\Gamma_{21}L\left\{2(c_{12}c_{23}s_{12})^2 - (2\alpha\Delta_{31}s_{12}c_{12}c_{23})^2 - 2(s_{13}c_{12}s_{12})^2 + 2s_{13}^2c_{12}^2s_{23}^2\cos(2\Delta_{31})\right\}\\ \nonumber
 &+\Gamma_{21}L\left\{ \alpha\Delta_{31} s_{13}\sin{\delta_{\rm CP}}\sin{2\theta_{12}}\sin{2\theta_{23}} + \frac{1}{2}s_{13}\sin{2\theta_{12}}\sin{2\theta_{23}}(\cos{(\delta_{\rm CP}+\Delta_{31}) + \cos{\delta_{\rm CP}}\cos{2\theta_{12}}})\right\} \\ \nonumber
 &+\Gamma_{32}L \left\{2s_{13}^2s_{23}^2\cos(2\Delta_{31}) - \alpha\Delta_{31}s_{13}\sin(\delta_{\rm CP}+2\Delta_{31})\sin{2\theta_{12}}\sin{2\theta_{23}}\right\}+ \\
 &-\frac{1}{2}s_{13}\sqrt{\Gamma_{21}\Gamma_{32}}L \left\{4s_{13}\cos{(2\Delta_{31})c_{12}^2s_{23}^2 + \cos{(\delta_{\rm CP} + 2\Delta_{31})\sin{2\theta_{12}\sin{2\theta_{23}}}}}\right\}, 
 \label{eq:appSManddeco}
\end{align}
where $c_{ij}$ and $s_{ij}$ are the cosines and sines of the mixing angles $\theta_{ij}$, respectively and $\Delta_{ij} = \frac{\Delta m^2_{ij}L}{4E}$, with $\Delta{m}^2_{ij}$ being the mass squared differences in vacuum. 
 Using appropriate assumptions, similar expressions can be obtained from Ref. \cite{Blennow:2005yk} where the authors have considered various damping scenarios.
The main feature of the decoherence correction is that $\Gamma_{21}$ affects the oscillation probability more than $\Gamma_{32}$. Retaining all linear terms in $\Gamma_{ij}$, the appearance probability reads:
\begin{eqnarray}\label{eq:appleading}
    P_{\mu e}^{deco}&\sim&2(\Gamma_{21}L) (c_{12}c_{23}s_{12})^2\\ \nonumber
    &&+\Gamma_{32}L \left\{2s_{13}^2s_{23}^2\cos(2\Delta_{31}) - \alpha\Delta_{31}s_{13}\sin(\delta_{\rm CP}+2\Delta_{31})\sin{2\theta_{12}}\sin{2\theta_{23}}\right\}
\end{eqnarray}
from which we observe that the first term is not suppressed by any of the small mixing angles.
This eq. \ref{eq:appleading} can then be used to understand the order of magnitude of $\Gamma_{21}$ and $\Gamma_{32}$ for which the decoherence term becomes the leading one. 
To this aim, we compare the leading SM probability contribution $4s_{13}^2s_{23}^2\sin^2\Delta_{31}$ with the leading term in $\Gamma_{21}$ and \gatm as shown in eq. \ref{eq:appleading}. In order for the decoherence correction to be larger than the leading SM term, the decoherence parameters must satisfy, at the oscillation maxima and at the best-fit values for the mixing angles, the following relations:
\begin{eqnarray}
    \Gamma_{21}L>\frac{2s_{13}^2s_{23}^2}{c_{12}^2s_{12}^2c_{23}^2}\sim 0.2
\end{eqnarray}
\begin{eqnarray}
    \Gamma_{32}L\gtrsim 2\,.
\end{eqnarray}

Note that $L = 360$ km for the \ess experiment. In particular, the second relation shows that we go beyond the validity of our perturbative expansion in $\Gamma_{32}$; we interpret this as a sign that, for the decoherence correction to be dominant over the standard one,  $\Gamma_{32}\gtrsim  10^{-22}$ GeV (while for the other parameter it is enough to fulfill $\Gamma_{21}\gtrsim  10^{-23}$ GeV).
 The other  crucial feature of the appearance probability is that both $\Gamma_{21}$ and $\Gamma_{32}$ may interfere with the measurement of $\delta_{\rm CP}$. 
We will explore the effect of decoherence in the PMNS matrix CP violating phase measurement at ESSnuSB in Sec. \ref{sec:CPV}.\\
In the disappearance channel, the oscillation probability reads
\begin{align}
\nonumber P_{\mu \mu}^{SM}&= 1 - \sin^2{2\theta_{23}}\sin^2{\Delta_{31}} + \alpha\Delta_{31}\sin{2\Delta_{31}}\sin{2\theta_{23}}\left(c_{12}^2 - 2s_{13}\cos{\delta}\sin{2\theta_{12}} s^2_{23}\right)\\ \nonumber
& - \left( 2\alpha\Delta_{31}c_{12}c_{23}\right)^2\left(c_{23}^2s_{12}^2 + s_{23}^2\cos{2\Delta_{31}}\right) + \left(2s_{13}s_{23}\sin{\Delta_{31}}\right)^2\cos{2\theta_{23}}\,,\\ \nonumber
P_{\mu \mu}^{deco}&= \Gamma_{21}L\Big[-2(c_{12}s_{12}c_{23})^2 + (\alpha\Delta_{31}\sin{2\theta_{12}}c_{23}^2)^2 + 2s_{13}\cos{\delta}s_{23}c_{23}^3(s_{12}c_{12} - \sin{3\theta_{12}}) \\ \nonumber
&- 2(s_{13}s_{23}c_{23})^2(s_{12}^4 +c_{12}^4) + s_{13}^2s_{23}^2\cos{2\Delta_{31}}(s_{12}^2+s_{12}^2\cos{2\theta_{23}}-2c_{12}^2s_{23}^2) -2\cos{2\Delta_{31}(s_{12}s_{23}c_{23})^2} \\ \nonumber
& + 8(s_{13}\cos{\delta}s_{12}c_{12}s_{23}c_{23}) - 4s_{13}\cos{\delta}\cos{2\Delta_{31}}s_{12}c_{12}s_{23}^3c_{23}\Big]\\ \nonumber
& + 2s_{23}^2\Gamma_{32}L \Big[\cos{2\Delta_{31}\left(-c_{23}^2 + 2\alpha^2\Delta_{31}^2c_{23}^2c_{12}^2 + s_{13}^2\cos{2\theta_{23}}\right)} \\ \nonumber
&+\alpha\Delta_{31}\sin{2\Delta_{31}\left(-2c_{12}^2c_{23}^2 + s_{13}\cos{\delta}\sin{2\theta_{12}\sin{2\theta_{23}}}\right)}
\Big]\\
& + s_{23}^2\cos{2\Delta_{31}}\sqrt{\Gamma_{21}\Gamma_{32}}L \Big[-2c_{23}^2s_{12}^2(-1+s_{13}^2) + s_{13}(2s_{13}c_{12}^2s_{23}^2 + \cos{\delta}\sin{2\theta_{12}\sin{2\theta_{23}}})
\Big]\,.
\label{eq: pmmdeco}
\end{align}
It can be observed that the channel is equally sensitive to both parameters, appearing not suppressed by any small parameters. In fact, the leading terms of the decoherence correction read:
\begin{eqnarray}
    P_{\mu \mu}^{deco}&\sim&-2(\Gamma_{21}L) \,c_{23}^2s_{12}^2(c_{12}^2c_{23}^2+s_{23}^2\cos{2\Delta_{31}})\\ \nonumber
    &&-2(\Gamma_{32}L) \, c_{23}^2s_{23}^2\cos{2\Delta_{31}}.
    \label{eq:decoleadingdis}
\end{eqnarray}
While one of the $\Gamma_{21}$ correction contains a term which does not depend on the atmospheric oscillation frequency $\Delta_{31}$, $\Gamma_{32}$ is proportional to $\cos{2\Delta_{31}}$, which is $\pm1$ at both oscillation maxima and minima.
We can again investigate for which values of the decoherence parameters the leading correction becomes comparable with the leading term $1-\sin^22\theta_{23}\sin^2\Delta_{31}$. In this case, at the oscillation maxima (which correspond to the disappearance probability minima), namely for $\Delta_{31}=(2n+1)\pi/2$, we obtain:
\begin{eqnarray}
    \Gamma_{21}(GeV)&\gtrsim&\frac{8\times10^{-20}}{L(\rm km)} \\ \nonumber
    \Gamma_{32}(GeV)&\gtrsim&\frac{4\times10^{-21}}{L(\rm km)}
\end{eqnarray}
which for correspond to $\Gamma_{21}\gtrsim2\times10^{-22}$ GeV and $\Gamma_{32}\gtrsim\times10^{-23}$ GeV at the ESSnuSB baseline of $360$ km.
Given these values, along with the ones obtained in the appearance channel, we can conclude that the appearance channel will dominate the sensitivity to $\Gamma_{21}$ while the disappearance channel will dominate the sensitivity to $\Gamma_{32}$, even though both channels will contribute in constraining the two parameters. 

\section{\ess experiment and simulation details}

In order to generate the probability, the event spectrum and to perform the sensitivity studies of \ess in the presence of decoherence, we make use of publicly available software GLoBES~\cite{Huber:2004ka, Huber:2007ji}. We have modified the probability engine to include decoherence as new physics
 effects and then performed the numerical computations to obtain the event rates and $\chi^2$. All the experimental details of \ess used in the present analyses are exactly the same as given in the CDR \cite{Alekou:2022emd} and have been incorporated in GLoBES. We consider a water Cherenkov far detector of fiducial volume 538 kt, 
located at a distance of 360 km at Zinkgruvan mine from the neutrino source in Lund. A powerful linear accelerator (linac) will be used to produce $2.7\times 10^{23}$ protons on target per year with a beam power of 5 MW and proton kinetic energy equal to 2.5 GeV. The updated neutrino fluxes with peak value around 0.25 GeV and event selection in the form of updated migration matrices have been adopted ~\cite{Alekou:2022emd}. The energy range [0, 2.5] GeV has been divided into 50 bins for the event calculation. In our analyses, we include both appearance ($\nu_\mu\to\nu_e$) and disappearance ($\nu_\mu\to\nu_\mu$) channels with their CP-conjugate transition, all equipped with the relevant backgrounds. We have considered a $5\%$ systematic errors for signal and $10\%$ systematic errors for backgrounds, unless otherwise mentioned. A total exposure of 10 years on the far detector is assumed (5 years run-time for neutrino beam and 5 years for antineutrino beam). 

\section{Understanding the decoherence at probability and event levels}
 \label{sec:prob_event_plots}
 
In this section
we first present a discussion on the appearance and disappearance probabilities to understand the effect of decoherence in neutrino oscillation at \ess energies. Then we study the total number of expected events in the presence of decoherence. Unless otherwise specified, the best-fit values of the standard oscillation parameters are adopted from NuFIT 5.2 (2022)~\cite{Esteban:2020cvm}, including Super-K atmospheric data and are listed in Table \ref{tab:t1}. Since recent global fits show a preference towards normal mass ordering (NO) for neutrinos \cite{Esteban:2018azc, deSalas:2020pgw, Capozzi_2020}, we present all our results considering NO only, i.e., for $\Delta m^2_{31} > 0$. However, we expect similar results when the neutrino mass ordering is inverted.  

\begin{center}
\begin{table}
\centering
\begin{tabular}{|c |c| c|} 
 \hline
Oscillation parameters ($3\nu$) & Normal ordering (NO) \\ [0.5ex] 
 \hline\hline
$\theta_{12}^{\circ}$ & $33.41^{+0.75}_{-0.72}$\\
 \hline
$\theta_{23}^{\circ}$ & $42.2^{+1.1}_{-0.9}$\\
\hline
$\theta_{13}^
{\circ}$ & $8.58^{+0.11}_{-0.11}$\\
\hline
$\delta_{CP}^{\circ}$ & $232^{+36}_{-26}$ \\
\hline
$\Delta m_{21}^2$ (eV$^2$) & $7.41^{+0.21}_{-0.20}\times 10^{-5}$ \\
\hline
$\Delta m_{31}^2$ (eV$^2$) & $+2.507^{+0.026}_{-0.027}\times 10^{-3}$ \\
\hline
\end{tabular}
\caption{ The best-fit value of the oscillation parameters in the standard three-flavour scenario. The values and their $1\sigma$ uncertainty intervals used in our calculations are taken from Ref.~\cite{Esteban:2020cvm}.}    \label{tab:t1}
\end{table}
\end{center}

\subsection{Discussion at the probability level}

\begin{figure}[H] 
\hspace*{-1cm}
     \centering
     \begin{subfigure}[b]{0.5\textwidth}
         \centering
         \includegraphics[width=8.5cm, height = 7cm]{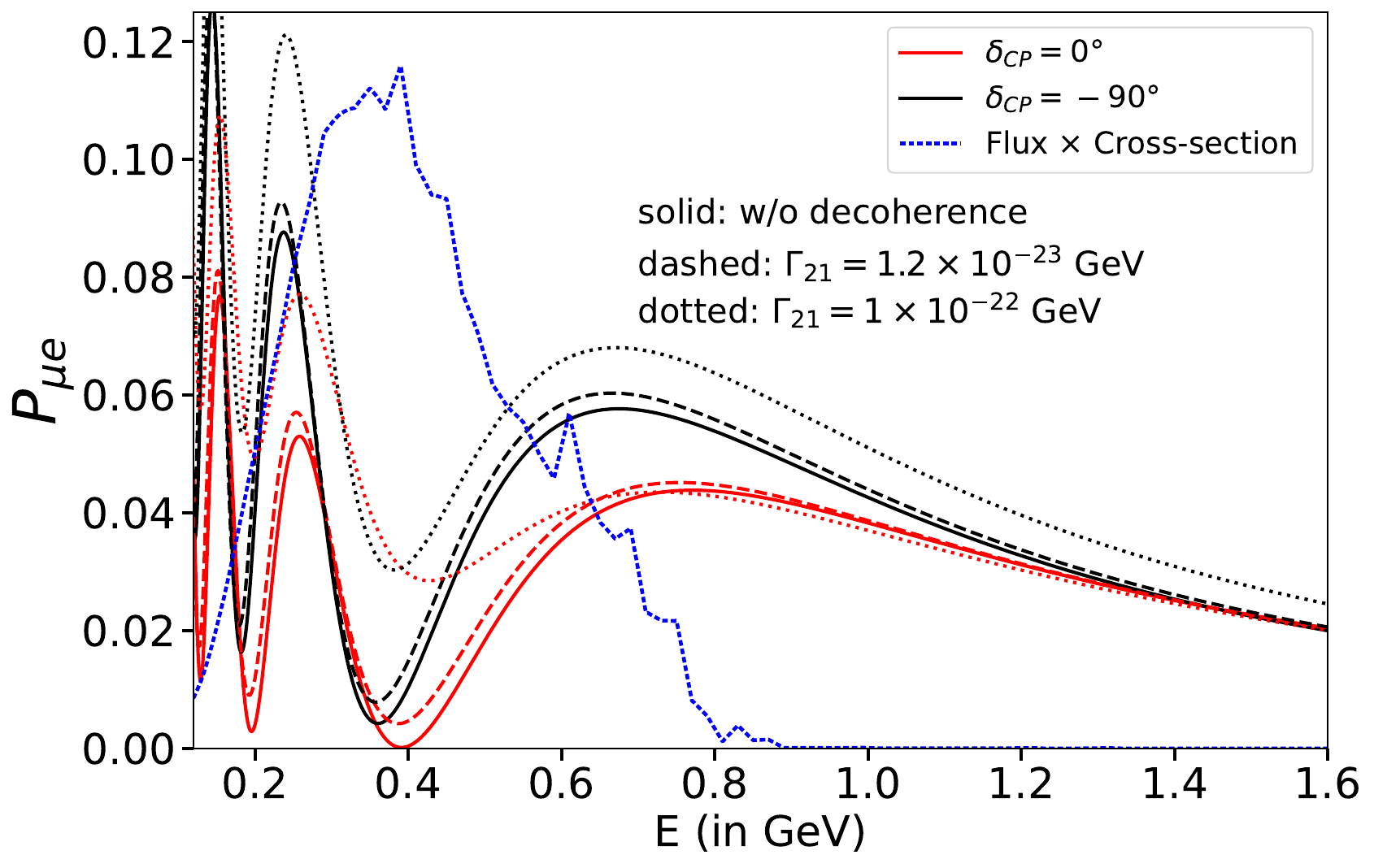}
         \caption{}
         
     \end{subfigure}
     \hfill
     \begin{subfigure}[b]{0.5\textwidth}
     \centering
     \includegraphics[width=8.5cm,height = 7cm]{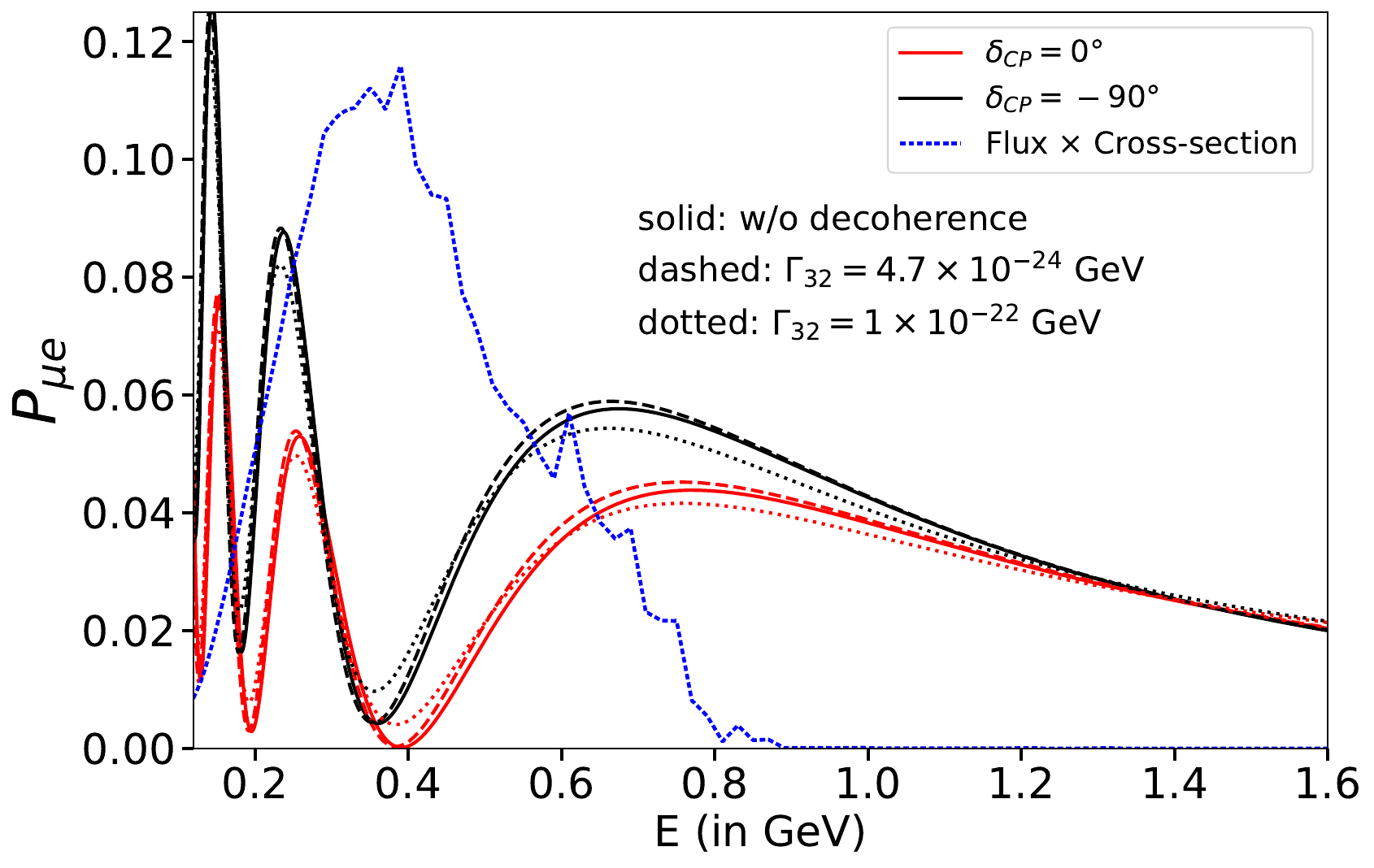}
     \caption{}
    
     \end{subfigure}
     \hfill
     \hspace*{-1cm}
    \begin{subfigure}[b]{0.5\textwidth}
         \centering
         \includegraphics[width=8.5cm,height = 7cm]{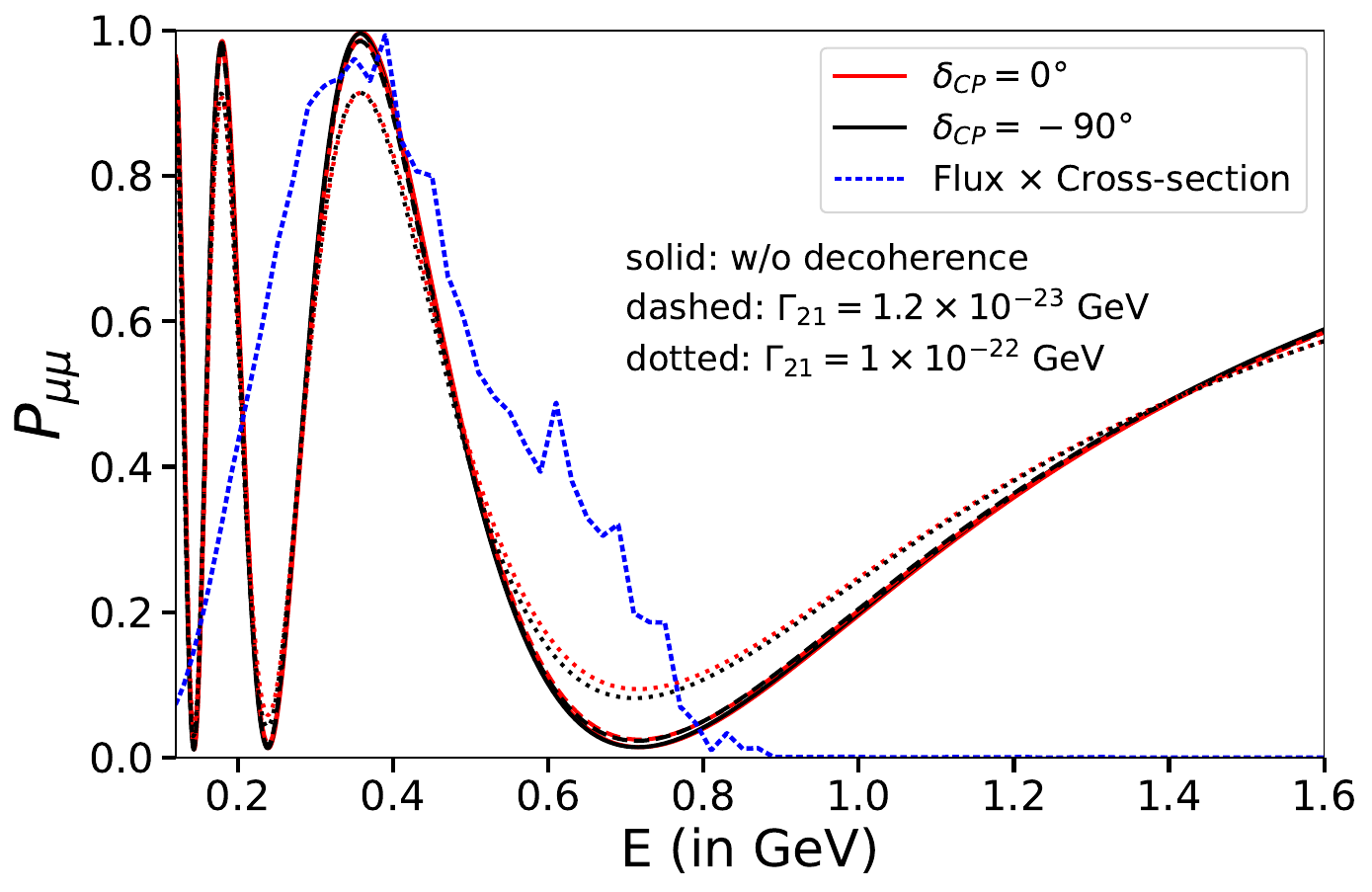}
         \caption{}
     \end{subfigure}
     \hfill
     \begin{subfigure}[b]{0.5\textwidth}
         \centering
         \includegraphics[width=8.5cm,height = 7cm]{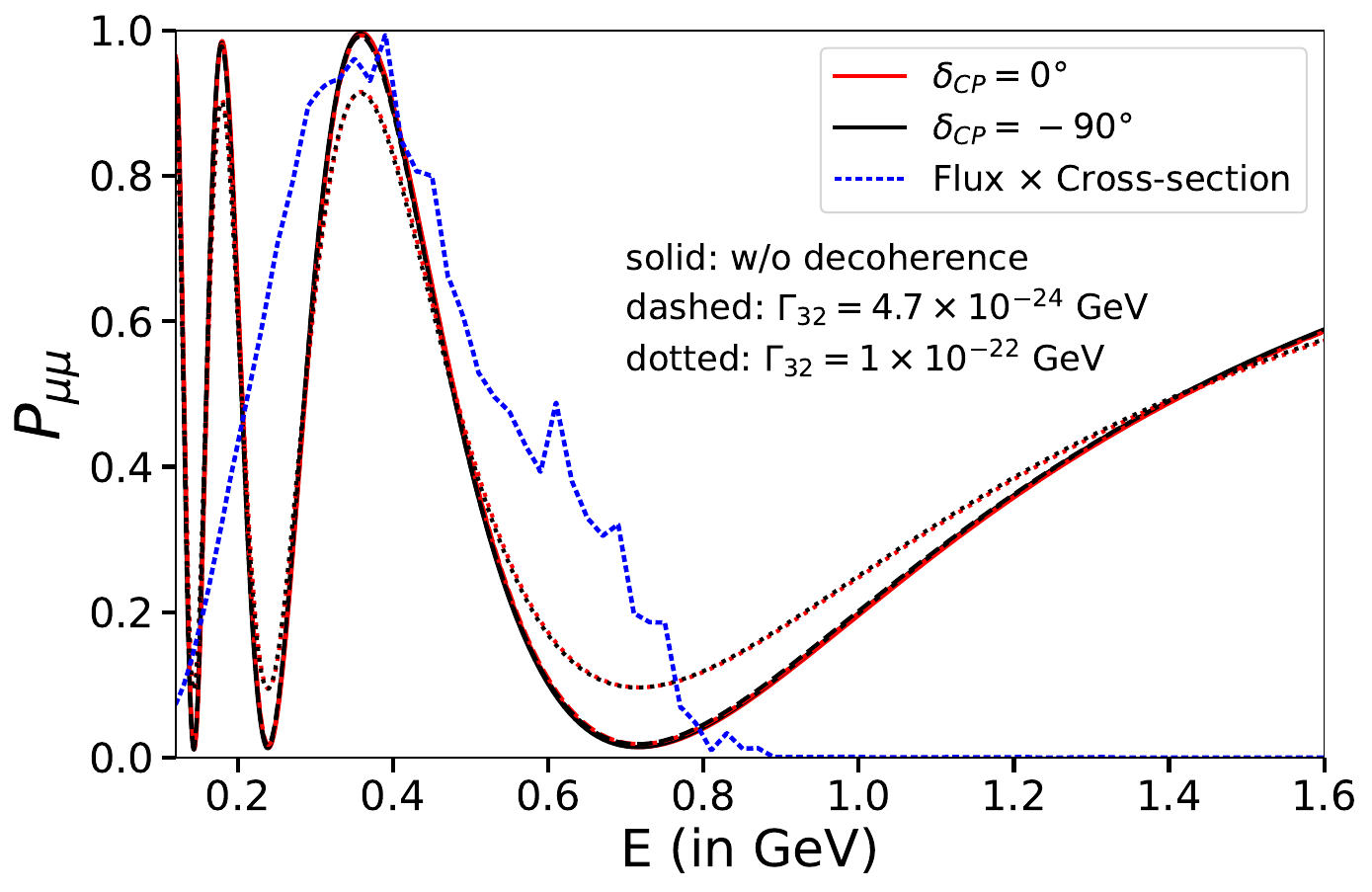}
         \caption{}
     \end{subfigure}
     \hfill
     \centering
     \caption{Appearance (top panel) and disappearance (bottom panel) neutrino oscillation probabilities as a function of neutrino energy for the baseline $ L = 360$ km. The left (right) panels are for the $\Gamma_{21}$ ($\Gamma_{32}$) case. }
        \label{fig:prob}
\end{figure}

In Fig. \ref{fig:prob}, we show neutrino oscillation probabilities as a function of neutrino energy relevant for the \ess experiment in the presence of decoherence. In this case, we consider the full oscillation probabilities in matter computed numerically. The top (bottom) panel is for the appearance (disappearance) channels. The left (right) panel depicts the effect of $\Gamma_{21}$ ($\Gamma_{32}$). In each panel, the solid curves refer to the SM probabilities, while the dotted and dashed curves are computed using two benchmark values of the decoherence parameters. 
The representative values of $\Gamma_{21}$ ($\Gamma_{32}$) have been fixed at the DUNE $~90\% \rm \rm ~C.L.$ limit \cite{BalieiroGomes:2018gtd} $1.2 \times 10^{-23}$ GeV ($4.7 \times 10^{-24}$) GeV and, for illustrative purposes for both $\Gamma$'s, at a second larger value, $1 \times 10^{-22}$ GeV. Moreover, two extreme values for $\delta_{\rm CP}$ have been chosen, corresponding to the case of maximal CP violation ($\delta_{\rm CP}=-90^\circ$, black curve) and vanishing CP violation ($\delta_{\rm CP}=0^\circ$, red curve). To show the energy region relevant for the ESSnuSB experiment, in each figure we also superimpose the \ess flux multiplied by the charged current (CC) neutrino cross-section. As already discussed in Sec.\ref{sec:formal}, the appearance probability $P_{\mu e}$ mostly depends on $\Gamma_{21}$; the effect of $\Gamma_{32}$ is small and can be seen mostly around the first oscillation maximum. The parameter $\Gamma_{21}$ is responsible for an increase of the probability at the first and second oscillation maxima as well as at the first oscillation minimum due to the positive sign of the leading decoherence correction. Moreover, the effect of $\Gamma_{21}$ is larger when the CP violation is maximum due to the probability terms which include the $\delta_{\rm CP}$ phase. The disappearance channel, on the other hand, almost equally depends on both decoherence parameters, with a slightly bigger sensitivity to the \gatm value. The minus sign in front of both leading corrections in $\Gamma_{21}$ and $\Gamma_{32}$ leads to smaller (larger) oscillation probabilities when $P_{\mu\mu}$ is maximum (minimum).

\newpage
\subsection{Discussion at the event level}

\begin{figure}[H]
     \centering
 \begin{subfigure}[b]{1\textwidth}
         \centering
         \includegraphics[width=15.0cm,height=7.5cm]{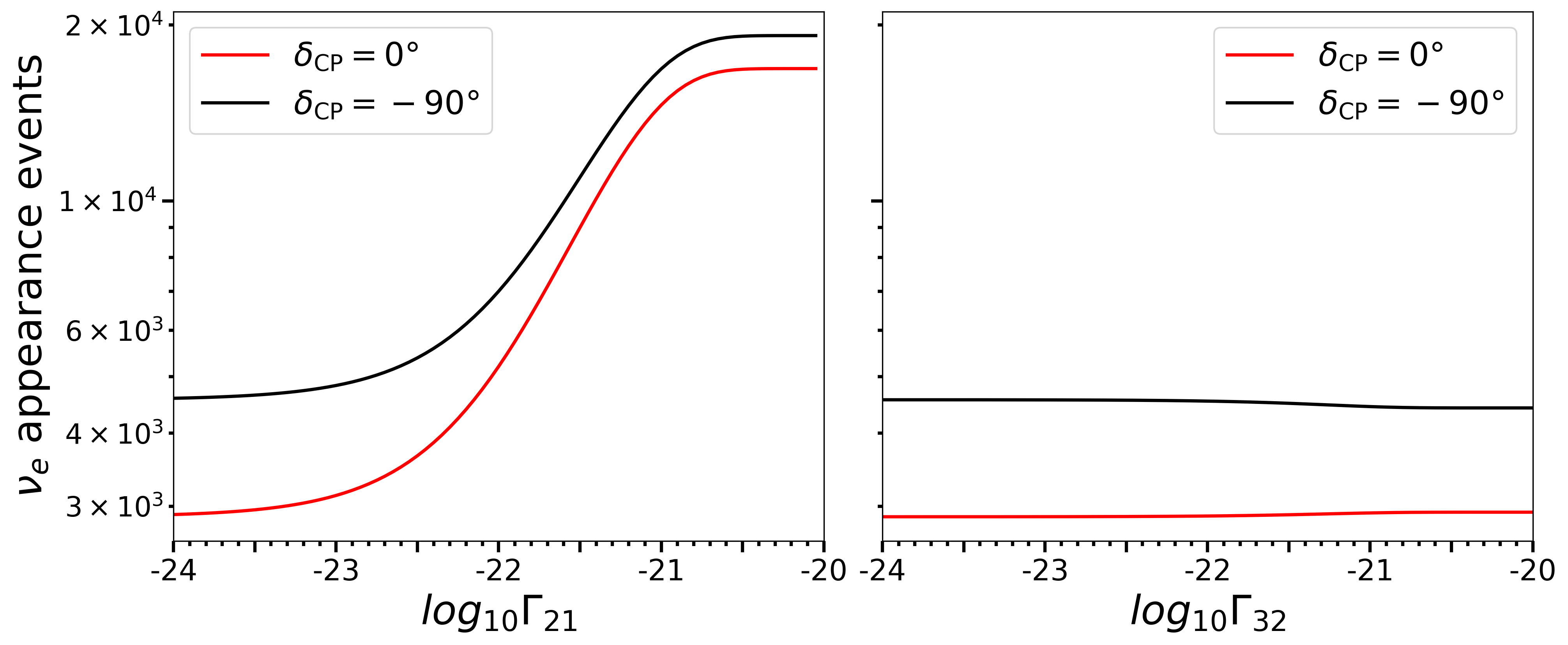}
         \caption{}
     \end{subfigure}
     \hfill
     \begin{subfigure}[b]{1\textwidth}
         \centering
         \includegraphics[width=15.0cm,height=7.5cm]{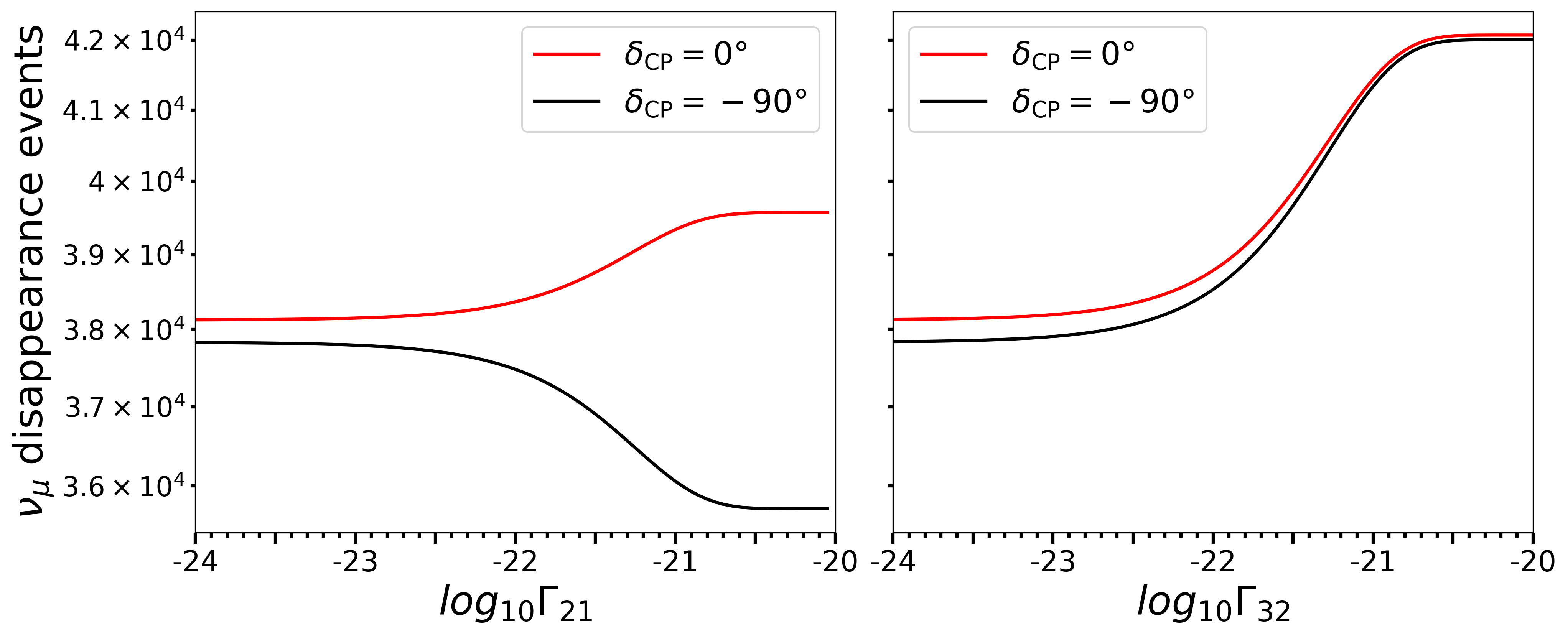}
         \caption{}
     \end{subfigure}
     \hfill
     \centering
\caption{Total number of events as a function of the decoherence parameters $\Gamma_{21}$ and $\Gamma_{32}$ and for two choices of $\delta_{\rm CP}$, $0^\circ$ and $-90^\circ$. }
        \label{fig:event360}
\end{figure}

In order to get an initial guess about the limits that \ess would set on decoherence parameters, we plot the total number of appearance (and disappearance) events as a function of the decoherence parameter for 10 years of running, 5 in neutrino and 5 in antineutrino mode.  The results are furnished in Fig. \ref{fig:event360} where black curves depict the case of maximal CP violation ($\delta_{\rm CP}=-90^\circ$) while red curves refer to the case of CP conservation ($\delta_{\rm CP}=0^\circ$). All the features discussed in Sec. \ref{sec:formal} can be appreciated in these plots. Indeed, in each case we can observe a transition between the SM dominated case and the decoherence dominated case. The transition begins for the values of \gsol and \gatm for which the decoherence correction overcomes the SM probability. Thus, \gsol becomes dominant for smaller values in the appearance channel than in the disappearance channel; moreover, \gatm does not affect in a relevant way the appearance channel. \\
For \gsol in appearance and \gatm in disappearance (case 1), the number of events increases drastically for both values of $\delta_{\rm CP}$  as $\Gamma$s get larger. In the other two cases (\gsol in disappearance and \gatm in appearance, case 2) we have an increment (decrement) of events for large values of the decoherence parameters when $\delta_{\rm CP}=0^\circ$ ($\delta_{\rm CP}=-90^\circ$). However, it can be noticed that for these choices of parameters, the variation of the total number of events (in case 2) when the decoherence parameters increase is substantially smaller than in case 1 and may also be affected by matter effects which we have not included in our analytical treatment. Since, as pointed out in Sec. \ref{sec:formal}, the main contribution to the \gsol ($\Gamma_{32}$) sensitivity will come from the appearance (disappearance) case, we will not discuss any further scenarios related to case 2.\\
Let us now discuss the behaviour of the number of events in the two most relevant frameworks, when \gsol increases in the appearance channel and when \gatm increases in the disappearance channel. In order to understand the effect of very large decoherence effects, we can compute the limit $\Gamma_{ij}\to \infty$ of eq. \ref{eq:finalprob}. In the electron appearance case, for $\Gamma_{21}\to\infty$ (notice that $\Gamma_{31}$ also tends to infinity in this case) and $\Gamma_{32}=0$ and expanding up to the first order in the small $\alpha$ and $\theta_{13}$, we obtain:
\begin{eqnarray}
    P_{\mu e}^{(\Gamma_{21}\to\infty)}&=&2 c_{12}^2s_{12}^2c_{23}^2\\ \nonumber
     & &+\frac{1}{2}s_{13}\sin2\theta_{12}\sin2\theta_{13}(\cos(\delta+2\Delta_{31})+\cos2\theta_{12}\cos\delta).
\label{eq:gsolinfty}
\end{eqnarray}
It is clear that this probability is always larger than the SM one (see the first line in eq. \ref{eq:appSManddeco}), 
because the leading term is second order in $\alpha$ and $\theta_{13}$. Thus, for any value of $\delta_{\rm CP}$, we expect in the decoherence dominated case a larger number of events with respect to the SM case. In the disappearance channel, taking the limit  $\Gamma_{32}\to\infty$ ($\Gamma_{31}\to\infty$) and $\Gamma_{21}=0$, we obtain at leading order 
\begin{eqnarray}
\label{eq:diff}
    P_{\mu \mu}^{(\Gamma_{32}\to\infty)}&=&\frac{1}{4}(3+\cos4\theta_{23})\,,
\end{eqnarray}
which does not depend on the atmospheric oscillation frequency $\Delta_{31}$. Taking the difference between \ref{eq:diff} and the leading term of the SM oscillation probability, we get:
\begin{eqnarray}
    P_{\mu\mu}^{SM}-P_{\mu \mu}^{(\Gamma_{32}\to\infty)}=\frac{1}{2}\sin^2 2\theta_{23}\cos2\Delta_{31} \, ,
\end{eqnarray}
which, at the oscillation maxima where $\cos2\Delta_{31}=-1$, suggests that the number of events is larger than in the absence of decoherence and independent, at least at the considered perturbative order, from the value of $\delta_{\rm CP}$.
\\

\newpage
\section{Exclusion Plots for ESSnuSB}

\begin{figure}[H]
     \centering        \includegraphics[width=7.5cm,height=7.5cm]{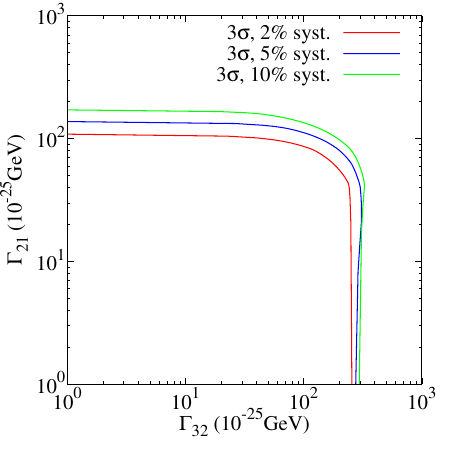}
        \includegraphics[width=7.5cm ,height=7.5cm]{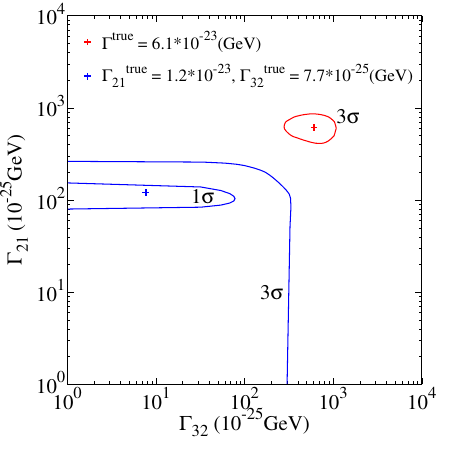}
\caption{\textit{Left}: $3\sigma$ contour plot (for 2 dof) for the decoherence parameters \gsol and \gatm in ESSnuSB. Different colours correspond to the different values of systematic uncertainties, as reported in the legend. \textit{Right}: The expected precision of ESSnuSB in measuring the decoherence parameters for 5\% systematics.   }
        \label{fig:gamma_CL}
\end{figure}

\begin{figure}[h]
\hspace*{-1.75cm}
\includegraphics[width=8cm,height=7.5cm]{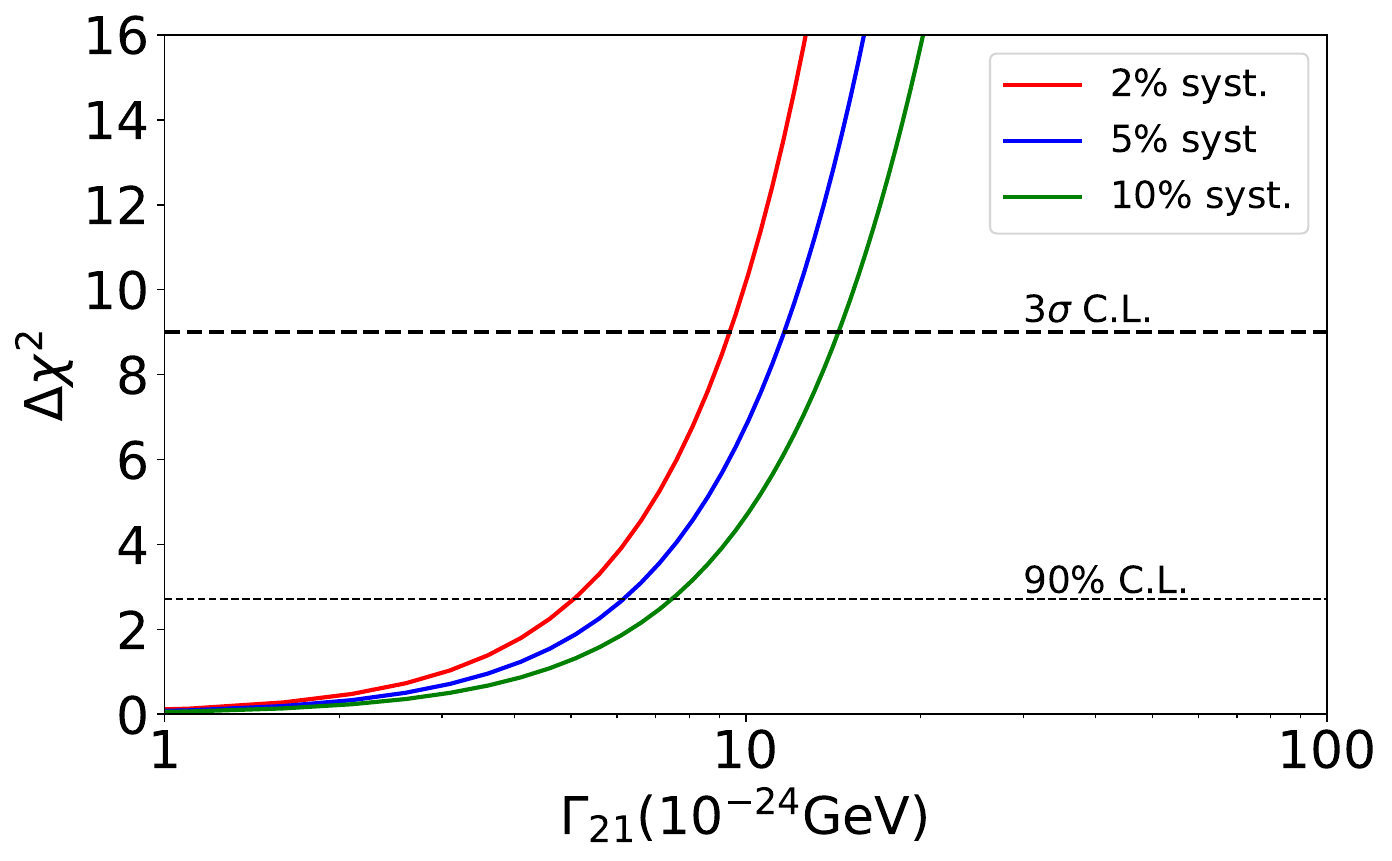}
\hspace*{0.5cm}
\includegraphics[width=8cm,height=7.5cm]{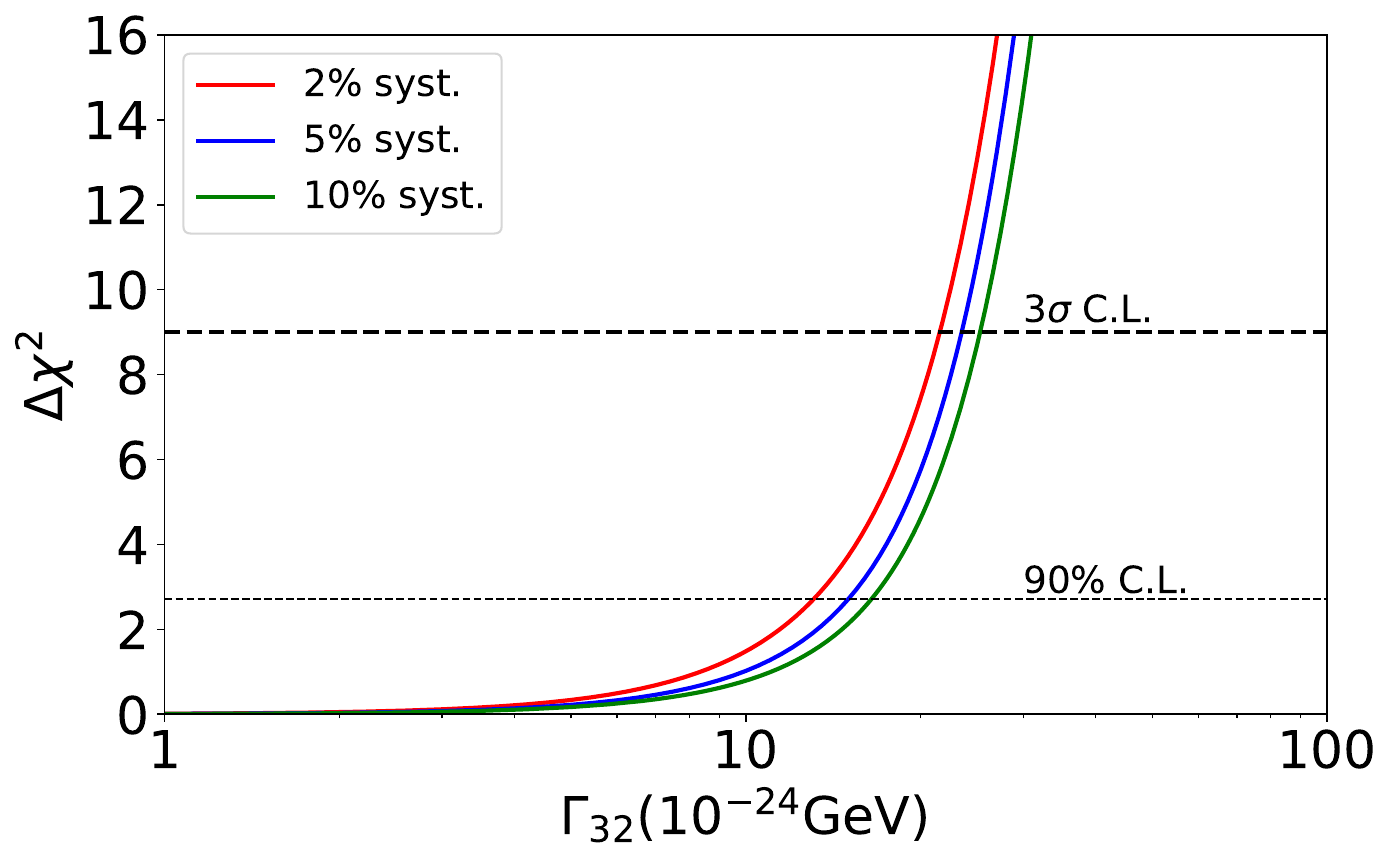}
\centering \includegraphics[width=8cm,height=7.5cm]{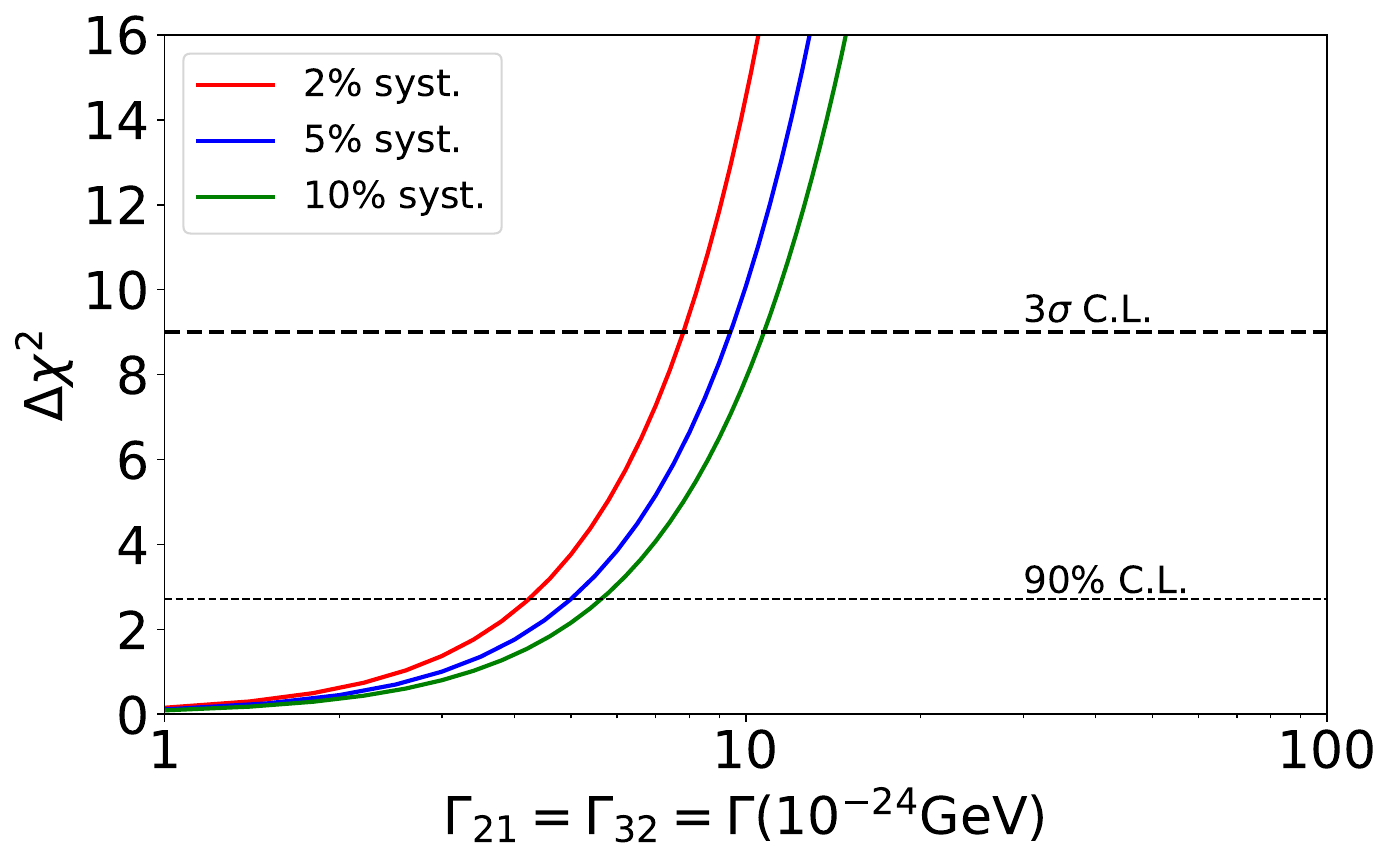}
\caption{Constraints on the decoherence parameters namely \gsol and \gatm from the ESSnuSB experiment. The upper-left (right) panel gives a sensitivity limit on \gsol ($\Gamma_{32}$). Different colours correspond to the different values of systematic uncertainties,
as reported in the legend. The dashed horizontal black lines represent the $90\%$ and $3\sigma$ levels. In the bottom plot we set $\Gamma_{21} = \Gamma_{32}$ in the test values.}
 \label{fig:fig:chisq}
\end{figure}

In this section, we want to explore the performances of the \ess experiment in constraining the decoherence parameters. As a first step, we check the correlations among the decoherence parameters. In particular, we show in Fig.~\ref{fig:gamma_CL} the 3$\sigma$ confidence level (C.L.), 2 degrees of freedom (dof) in the \gatm-\gsol plane. The analysis has been performed using a Poissonian $\chi^2$ defined as
\begin{equation}
    \chi^2(\Vec{\lambda},a)=\sum_{i=1}^n 2\left((1+a)T_i-O_i+O_i\log\frac{O_i}{(1+a)T_i}\right)+\frac{a^2}{\sigma_a^2}\,,
\end{equation}
where $\Vec{\lambda}$ is the set of oscillation parameters needed to compute the rates, $\sigma_a$ is the normalization error, $n$ is the number of energy bins, $O_i$ are the observed rates and $T_i$ are the theoretical rates used for the fit. The systematic uncertainties are treated with the \textit{pull-method} \cite{Huber:2002mx,Fogli:2002pt} implemented in GLoBES using the nuisance parameter $a$. In the left panel of Fig. \ref{fig:gamma_CL} we set as true values $\Gamma_{21}=\Gamma_{32}=0\,$ and then we fit with the decoherence hypothesis, marginalizing over all not shown oscillation parameters within the uncertainties reported in Tab. \ref{tab:t1}, except for the solar ones, which we held fixed to their central values. The CP violating phase $\delta_{\rm CP}$ is left free to vary in its $[0^\circ-360^\circ]$ range. We also show here the effects of systematics on the \gatm-\gsol correlation using three benchmark values, namely an optimistic 2\% (red curve), the standard \ess 5\% (blue curve) and a pessimist 10\% (green curve). It is clear that the two decoherence parameters are not correlated, the 3$\sigma$ limit on one of the two being independent of the test value of the other. This is clear from the oscillation probabilities (Eqs. \ref{eq:appSManddeco} and \ref{eq: pmmdeco}), where the only correlation comes from the small mixed $\sqrt{\Gamma_{21}\Gamma_{32}}$ term. It is also interesting to notice that the choice of systematics does not affect the analysis in a relevant way, especially for $\Gamma_{32}$. This result mainly comes from the fact that the decoherence parameters modify the oscillation probabilities exponentially. In the right panel of Fig. \ref{fig:gamma_CL}, we show the foreseen precision on the new physics parameters 
obtained with 5\% systematics for two benchmark true values of  $\Gamma_{ij}$; the first one inspired by the relatively large 90\% C.L. limits achieved by T2K+MINOS\footnote{ Note that the bound for T2K+MINOS is calculated under the assumption $\Gamma_{32}=\Gamma_{21} = \Gamma_{31}$ whereas in our formalism, $\Gamma_{32} = \Gamma_{21}$ will imply $\Gamma_{31} = 0$ (cf. eq. \ref{eq:gamma_relation}). Therefore, though the bound obtained by T2K+MINOS cannot be directly compared within our formalism, we have used their bound as a reference point to understand the precision of the decoherence parameters in case large decoherence exist in Nature.} \cite{Gomes:2020muc}, namely $\Gamma_{21}=\Gamma_{32}=6.1\times 10^{-23}$ GeV (red curve); the second one taken from the best possible 90\% C.L. limits achievable at DUNE, $\Gamma_{21}=1.2\times10^{-23}$ GeV and $\Gamma_{32}=7.7\times10^{-25}$ GeV (blue curve). Notice that the DUNE limits have been obtained in Ref.~\cite{BalieiroGomes:2018gtd} using the standard DUNE flux for \gsol and the high energy DUNE flux for $\Gamma_{32}$. These two benchmark choices allow us to observe two interesting results. When both decoherence parameters are large and lie in a region already excluded (see left panel), \ess is capable of obtaining a very precise measurement of the decoherence parameters. Indeed, the allowed region within the red curve is rather small and does not include the standard oscillation scenario $\Gamma_{21}=\Gamma_{32}\to0$. On the other hand, when the other benchmark values are taken into account, it is clear that at 3$\sigma$ \ess may not be able to exclude the standard oscillation scenario. However, if we consider only the 1$\sigma$ range, \gsol is measured with a very good uncertainty, while \gatm does not have a lower bound, exploiting the fact that the DUNE high energy flux may allow setting a limit on \gatm which is not reachable by \ess\footnote{The authors of Ref. \cite{BalieiroGomes:2018gtd} show that there exists a new matter effects resonance around 10 GeV for DUNE driven by \gatm at DUNE. Thus, a high energy flux may be extremely powerful in constraining this parameter. Such a resonance is not observable at \ess since it would require neutrinos with energy of $\sim3$ GeV.}. \\
After checking that the correlations between the two decoherence parameters are negligible, we proceed to consider in Fig. \ref{fig:fig:chisq} the sensitivity to \gsol when \gatm= 0 (top left panel), to \gatm when \gsol= 0 (top right panel) and in the case \gsol= \gatm (bottom panel). For this computation, we generate data in the hypothesis of no decoherence and we fit them using the probabilities in the presence of decoherence. The marginalization has been performed over all the oscillation parameters but the solar ones, as before. We also checked that marginalization on the not-shown decoherence parameter in the first two cases does not affect our results in a relevant way, confirming that the correlation among the parameters at \ess is negligible. We also show the results for different values of the normalization systematic uncertainty, namely 2\% (red curves), 5\% (blue curves) and 10\% (green curves). The 3$\sigma$ and 90\% C.L. bounds are summarized in Tab.\ref{tab:bound_table} for the standard 5\% systematics case along with the 2\% and 10\% systematics cases. The main results are that \ess in the nominal conditions (5\% systematics) may be able to set the two 90\% limits $\Gamma_{21}<6.15\times10^{-24}$ GeV and $\Gamma_{32}<1.50 \times 10^{-23}$ GeV. With the further constraint $\Gamma_{21}=\Gamma_{32}$, \ess is expected to set the limit $\Gamma_{21}=\Gamma_{32}<4.99\times10^{-24}$ GeV. Comparing this result with the existing best LBL constraint in eq. \ref{bounds:MINOS}, we observe that \ess performances may allow to overcome the MINOS/MINOS+ ones by roughly 40\%. However, solar and atmospheric neutrinos are expected to be more powerful in this context, being able to probe higher energy neutrinos which could encounter new matter resonances driven by decoherence parameters. The next generation experiment DUNE, on the other hand, is expected to obtain similar limits with its standard neutrino flux: $\Gamma_{21}<1.2 \times 10^{-23}$ GeV and $\Gamma_{32}<4.7\times10^{-24}$. The former is less stringent than the \ess one, while the latter is 5 times better than the \ess one, because of the higher energies reached by DUNE and because of the more pronounced matter effects. The role of the systematics in ESSnuSB, as already mentioned, is not crucial; however, if 2\% normalization uncertainty will be achieved, the limits $\Gamma_{21}<5.06\times10^{-24}$ GeV and $\Gamma_{32}<1.31 \times 10^{-23}$ GeV could be obtained, which mark an improvement by about a factor $2-3$ compared to the nominal case. In conclusion, \ess should be able to put very competitive bounds on the decoherence parameters with respect to the current and future LBL experiments. The complementarity among the limits obtained by accelerator, atmospheric and solar neutrinos may allow to further reduce the allowed parameter space of this model.


\begin{table}[H]
\hspace*{-1cm}
\begin{tabular}{|L|c|c|c|c|c|c|}
\hline
\multirow{3}{*}{}{Decoherence
Parameter} & \multicolumn{3}{c|}{$3\sigma$ C.L. (in GeV)} & %
    \multicolumn{3}{c|}{$90\%$ C.L. (in GeV)}\\
\cline{2-7}

&$2\%$ syst. & $5\%$ syst. & $10\%$ syst. & $2\%$ syst. & $5\%$ syst. & $10\%$ syst.\\
\hline
$\Gamma_{21}  (= \Gamma_{31})$ when $\Gamma_{32} = 0$& $0.94\times 10^{-23}$  & $1.16\times 10^{-23}$  & $1.44\times 10^{-23}$  &$5.06\times 10^{-24}$ &$6.15\times 10^{-24}$ &$7.45\times 10^{-24}$ \\
\hline
$\Gamma_{32}  (= \Gamma_{31})$  when $\Gamma_{21} = 0$&$2.16\times 10^{-23}$ &$2.35\times 10^{-23}$ &$2.53\times 10^{-23}$ &$1.31\times 10^{-23}$ &$1.50\times 10^{-23}$ &$1.64\times 10^{-23}$ \\
\hline
$\Gamma_{21}$ = $\Gamma_{32}$ $ (\Gamma_{31} = 0)$  &$7.81\times 10^{-24}$ & $9.41\times 10^{-24}$&$10.74\times 10^{-24}$ &$4.23\times 10^{-24}$ &$4.99\times 10^{-24}$ & $5.64\times 10^{-24}$\\
\hline
\end{tabular}
\caption{Constraints on decoherence parameters for $2\%, 5\%$ and $10\%$ systematics from ESSnuSB experiment. We also report   the effects of using three different values of systematics. 
}
\label{tab:bound_table}
\end{table}

\section{Correlations}
\label{sec:correlation}

\begin{figure}[H]
     \centering
         \includegraphics[width=7.5cm,height=7.5cm]{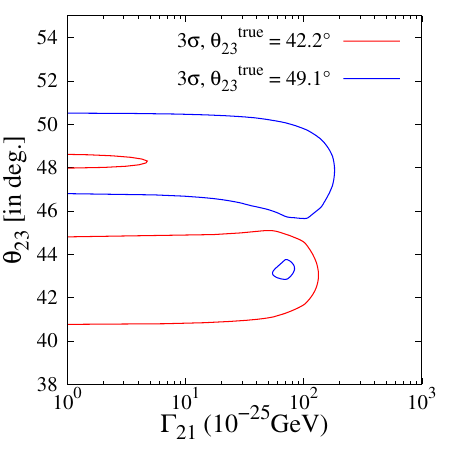}
        \includegraphics[width=7.5cm,height=7.5cm]{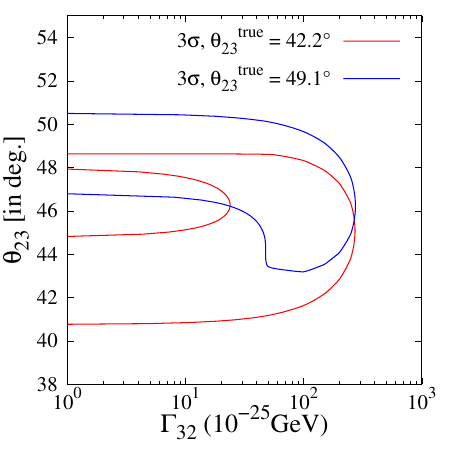}
\caption{Effect of decoherence on $\theta_{23}$. In both panels, two distinct true values for the atmospheric angles have been chosen, $\theta^{\rm true}_{23}=42.2^\circ, 49.1^\circ$. On the left plot, the $\Gamma_{21}-\theta_{23}$ correlation is shown, while in the right panel, we present the $\Gamma_{32}-\theta_{23}$ correlation.} 
        \label{fig:gamma_th23}
\end{figure}

\begin{figure}[H]
     \centering
         \includegraphics[width=7.5cm,height=7.5cm]{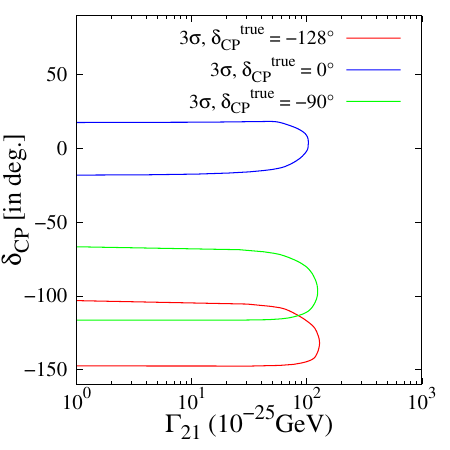}
        \includegraphics[width=7.5cm,height=7.5cm]{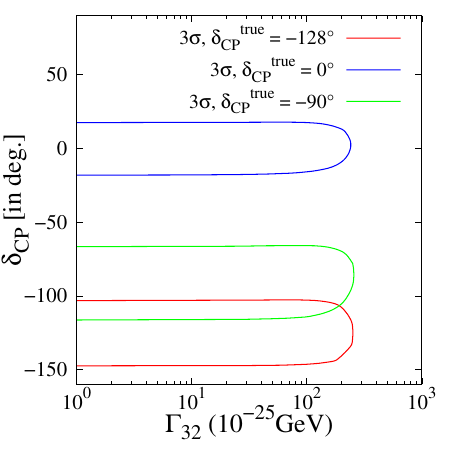}
\caption{Same as Fig. {\ref{fig:gamma_th23}}, but in the $\Gamma_{ij}-\delta_{\rm CP}$ planes.}
        \label{fig:gamma_dcp}
\end{figure}

In this section, we will explore the correlations among the decoherence parameters and the two not-well known standard oscillation parameters for ESSnuSB, namely $\delta_{\rm CP}$ and $\theta_{23}$. 
To perform this analysis, the events spectra have been produced under the assumption of no-decoherence and employing the best-fit values for the standard oscillation parameters, Tab.\ref{tab:t1}. The fit has been obtained by marginalizing over all the not-shown standard oscillation parameters except the solar ones.
In Fig. \ref{fig:gamma_th23}, we show the 3$\sigma$ allowed regions in the $\Gamma_{21}-\theta_{23}$ (left panel) and $\Gamma_{32}-\theta_{23}$ (right panel) planes for two different choices of the atmospheric mixing angle true values, one in the lower octant ($42.2^\circ$) and one in the upper octant ($49.1^\circ$). These correspond to the best-fits from \cite{Esteban:2020cvm} with and without SK atmospheric data. In the absence of decoherence, the \ess results might not be able to resolve the $\theta_{23}$ octant if the true value is $\theta_{23}=42.2^\circ$; indeed, there are always allowed values which lie in the upper octant. On the other hand, for $\theta_{23}=49.1^\circ$ and $\Gamma_{ij}\to0$, the octant degeneracy appears to be broken. In the presence of decoherence, we have two opposite behaviours. When \gsol increases, the octant ambiguity is solved and for both chosen true values of the atmospheric mixing angle, the $\theta_{23}$ octant might be resolved. This is because the leading decoherence correction in the appearance channel, which is the most sensitive to $\Gamma_{21}$, is proportional to $\cos^2\theta_{23}$ (see eq. \ref{eq:appleading}). On the other hand, when \gatm increases, the octant degeneracy is more pronounced for $\theta_{23}=42.2^\circ$ and appears also for $\theta_{23}=49.1^\circ$. This is clearly understood in eq. \ref{eq:decoleadingdis}, where we showed that the \gatm correction is proportional to $\sin^22\theta_{23}$. \\
In Fig. \ref{fig:gamma_dcp} we present the results in the \gsol - \dcp~(left panel) and \gatm - \dcp~(right panel) planes for three values of the CP violating phase corresponding to maximal CPV ($\delta_{CP}=-90^\circ$), no CPV ($\delta_{CP}=0^\circ$) and to the best-fit from \cite{Esteban:2020cvm}, namely $\delta_{CP}=-128^\circ$. In this case, we observe no relevant correlations between \dcp~and the two decoherence parameters. However, the effects of \gatm and \gsol on the \dcp~determination might become important if the decoherence parameters are large enough to overcome the \ess sensitivity, i.e. if their value will be measurable at the experiment. We will explore in details this topic in the next section.

\section{CPV sensitivity of ESSnuSB in Presence of Decoherence}
\label{sec:CPV}

\begin{figure}[h]
\centering
\includegraphics[width=14.2cm,height=9.2cm]{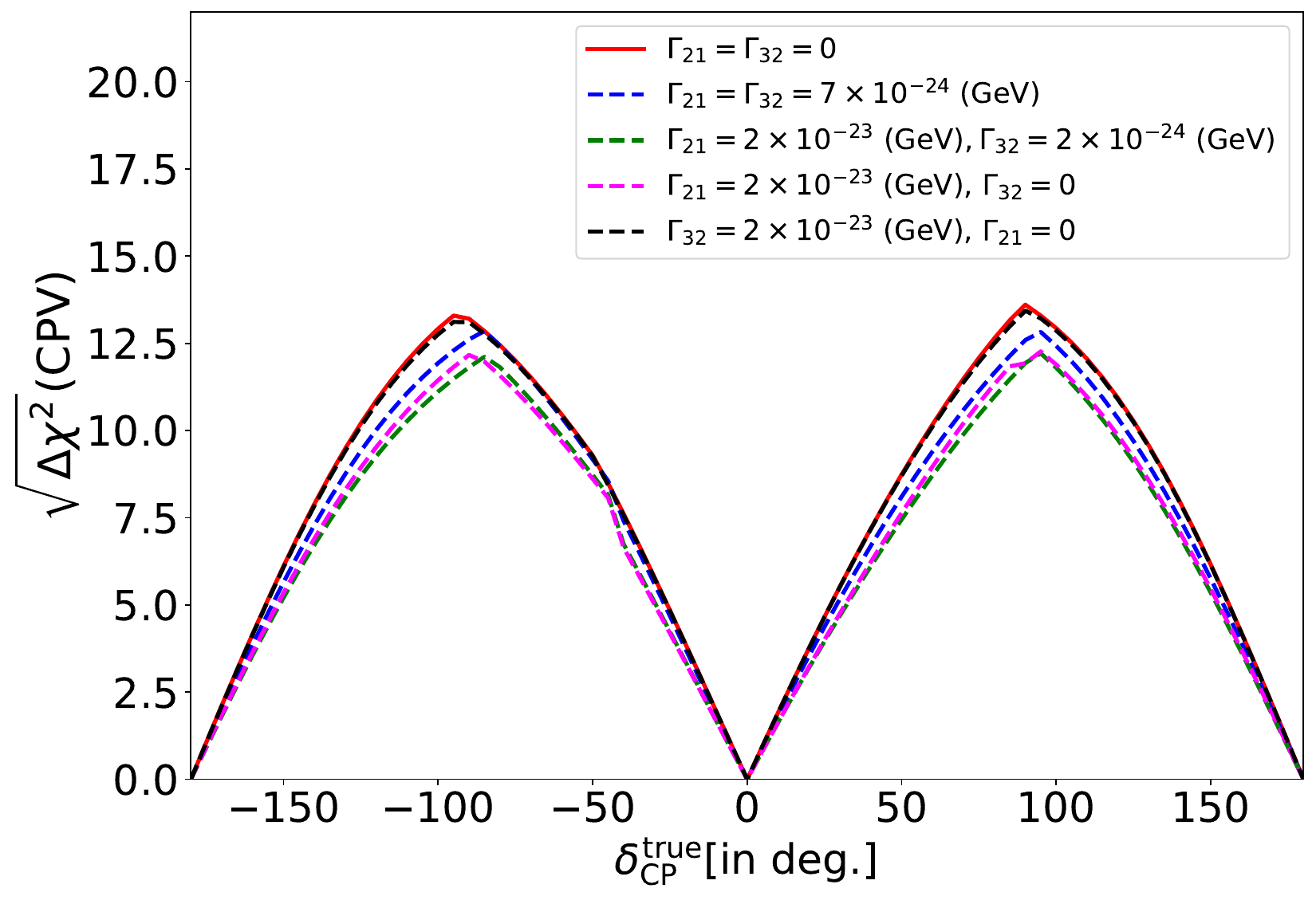}
\caption{CP violation sensitivity of \ess for different values of decoherence parameters. Here both true and test hypotheses assume decoherence. }
  \label{fig:cpv2}
\end{figure}

\begin{figure}[h]
\centering
\includegraphics[width=15cm,height=9cm]{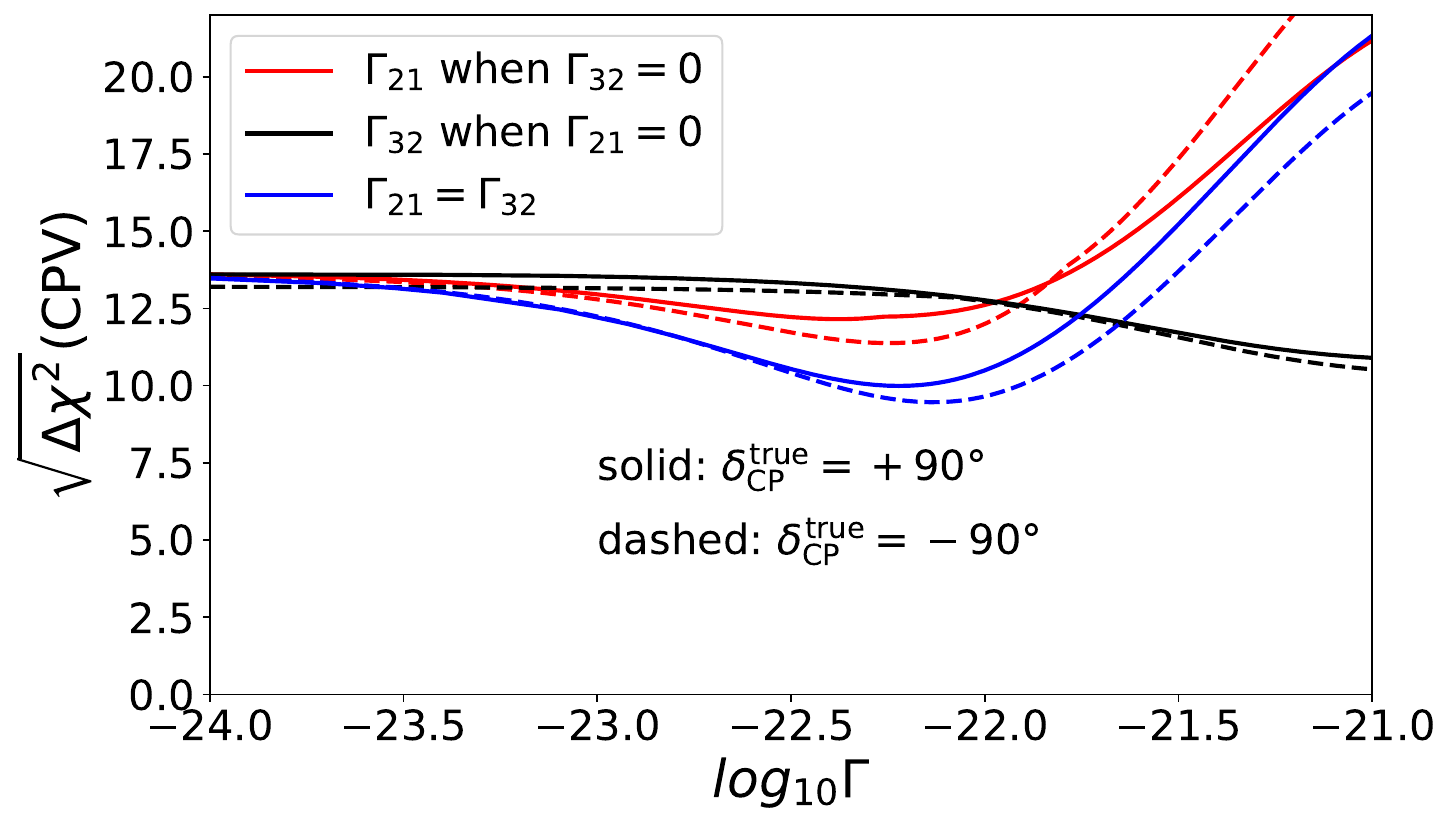}
\caption{CP violation sensitivity of ESSnuSB as a function of decoherence parameters ($\Gamma_{ij}$).}
 \label{fig:cpv_gamma}
\end{figure}

In this section, we will exploit the interplay between the decoherence parameters and the \ess CP violation sensitivity. This aspect is crucial since the main purpose of the experiment is the precise \dcp~measurement and it is essential to understand whether the presence of new physics could spoil its potential to do so. It has been shown that \ess should be capable of reaching 12.5$\sigma$ sensitivity for maximal CP violation and reach at least 5$\sigma$ sensitivity for roughly 75\% of the possible phase values \cite{Alekou:2022emd,ESSnuSB:2023lbg}.This outperforms the sensitivity of all next-generation LBL oscillation experiments \cite{Agarwalla:2022xdo}.
In the following, we will show whether the presence of decoherence could destroy such good prospects. In Fig. \ref{fig:cpv2} we plot the CPV sensitivity in units of $\sqrt{\Delta \chi^2}$, where
\begin{equation}
    \Delta\chi^2=\chi^2(\mathrm{deco},CPV)-\chi^2(\mathrm{deco},\delta_\mathrm{CP}=0,180^\circ) \, .
\end{equation}
Thus, we fix the same decoherence parameters in both true and fit values of the oscillation parameters. We show the results for several choices of \gsol and $\Gamma_{32}$.  For reference, we add the red curve which represents the sensitivity without decoherence. It is clear that, for $\Gamma\sim\mathcal{O}(10^{-[23;24]})\, \, \mathrm{GeV}$ which is the order of magnitude of the \ess bounds, the effect on the \dcp~sensitivity is limited. Even though a mild reduction at the largest of 15\% for the CPV sensitivity at the two maximal values $\delta_{\rm CP} = \pm90^\circ$ for $\Gamma_{21}=2\times10^{-23}$ GeV and $\Gamma_{32}=2\times10^{-24}$ GeV is observed, the \ess experiment remains extremely powerful in the context of the \dcp~measurement even with decoherence. It is interesting to notice that the sensitivity maxima slightly change their position when decoherence parameters are increased. \\
Let us now discuss in more detail the effects of \gsol and \gatm on the $\delta_{\rm CP}$ sensitivity at the maximal values. In Fig. \ref{fig:cpv_gamma} we show the \ess sensitivity for $\delta_{CP}=90^\circ$ (solid lines) and $\delta_{CP}=-90^\circ$ (dashed lines) as a function of \gsol and $\Gamma_{32}$, up to $10^{-21}$ GeV. In this case, in order to avoid numerical instabilities, in the $\Delta\chi^2$ computation we fixed $\theta_{23}$ to its best-fit value; however, we check that the marginalization procedure over $\theta_{23}$ had a negligible impact on the results.
We first discuss the \gsol case (red line). From eq.(\ref{eq:appSManddeco}) we obtain that the appearance terms containing $\delta_{\rm CP}$ reads:
\begin{eqnarray}
    P_{\mu e}^{\delta_{\mathrm{CP}}}&=&2\alpha\Delta_{31}s_{13}\sin2\theta_{23}\sin2\theta_{12}\cos(\delta+\Delta_{31})\sin\Delta_{31}\\\nonumber
    &&+\frac{1}{2}\Gamma_{21}L s_{13}\sin2\theta_{23}\sin2\theta_{12}(\cos(\delta+\Delta_{31})+\cos2\theta_{12}\cos\delta) \\\nonumber
    &&+ 2\alpha\Delta_{31}s_{13}c_{12}^3s_{12}\sin2\theta_{23}\sin\delta \, . 
    \label{eq:probgsol}
\end{eqnarray}
Taking into account only the term proportional to $\sin\delta$, which drives the sensitivity to the CP-violation, the maximal CP contribution ($\sin\delta=\pm1$) is: 
\begin{equation}
\label{eq:amue_g21}
    |P_{\mu e}^{\mathrm{CP-odd}}|\propto |2\alpha\Delta_{31}(\Gamma_{21}L-2\sin^2\Delta_{31})-\Gamma_{21}L\sin2\Delta_{31}| \, .
\end{equation}
Thus, for small values of $\Gamma_{21}$, the decoherence parameter acts here as a correction to the SM CP-odd term with the opposite sign, thus reducing its absolute value. For this reason, we expect the CP-violation sensitivity to first slightly decrease. Then, when the decoherence term becomes dominant, the relevant probability increases along with $\Gamma_{21}$, improving the sensitivity. 
This behaviour is compatible with the one shown in the red line of Fig. \ref{fig:cpv_gamma}.\\
When we consider $\Gamma_{32}$, instead, the situation is different. Indeed, as we already mentioned, the decoherence correction in the appearance channel is always as suppressed as the leading terms in the probabilities; thus, the new physics correction becomes dominant only for extremely large values of $\Gamma_{32}$, which break the expansions shown in Sec. \ref{sec:prob:vacuum}. Thus, we expect the \gatm effect on the CPV sensitivity to be small. From eq. \ref{eq:appSManddeco}, the terms depending on \dcp~are the following:
\begin{eqnarray}
    P_{\mu e}^{\delta_{\mathrm{CP}}}&=&2\alpha\Delta_{31}s_{13}\sin2\theta_{23}\sin2\theta_{12}\cos(\delta+\Delta_{31})\sin\Delta_{31}\\
    &&-\Gamma_{32}L\alpha\Delta_{31}s_{13}\sin2\theta_{23}\sin2\theta_{12}\sin(\delta+2\Delta_{31}) \, .
    \label{eq:probgatm}
\end{eqnarray}
For maximal CPV, we get:
\begin{equation}
   |P_{\mu e}^{\mathrm{CP-odd}}|\propto |\Gamma_{32}L\cos2\Delta_{31}+2\sin^2\Delta_{31}| \, ,
\end{equation}
where, clearly, the small decoherence contribution diminishes the probability around the oscillation maxima (where $\cos2\Delta_{31}\sim-1$) and therefore the CPV sensitivity when \gatm increases. This behaviour is confirmed by the black line in Fig. \ref{fig:cpv_gamma}. Being \gsol dominant with respect to \gatm in the appearance channel, the case in which \gatm=\gsol resembles the same feature as the case \gsol $\ne 0$, \gatm$=0$.

\section{Precision measurement of $\delta_{\rm CP}$ in Presence of Decoherence}
\label{sec:precision}
\begin{figure}[H]
  \hspace*{-1cm}       \includegraphics[width=9cm,height=8cm]{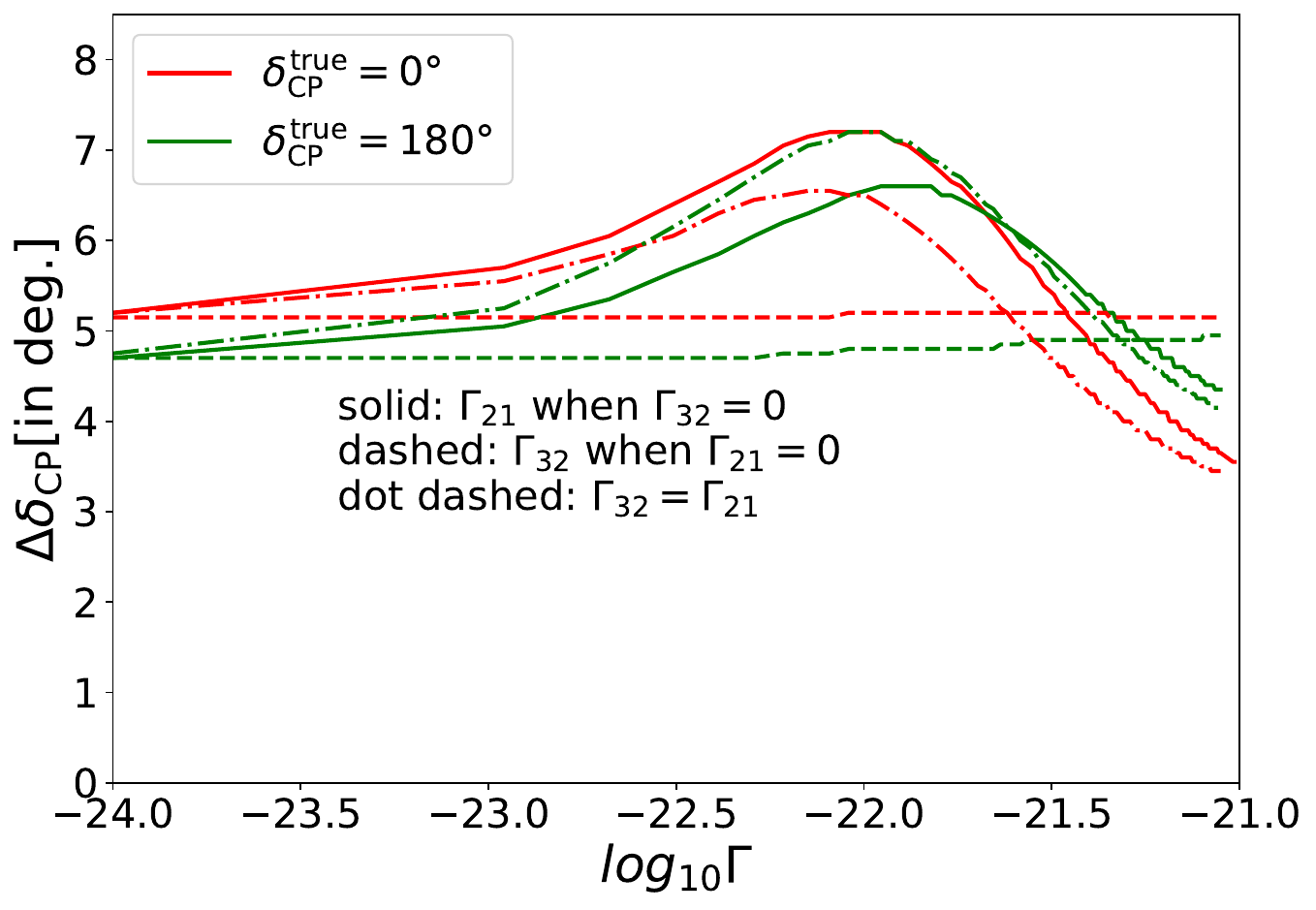}
        \includegraphics[width=9cm,height=8cm]{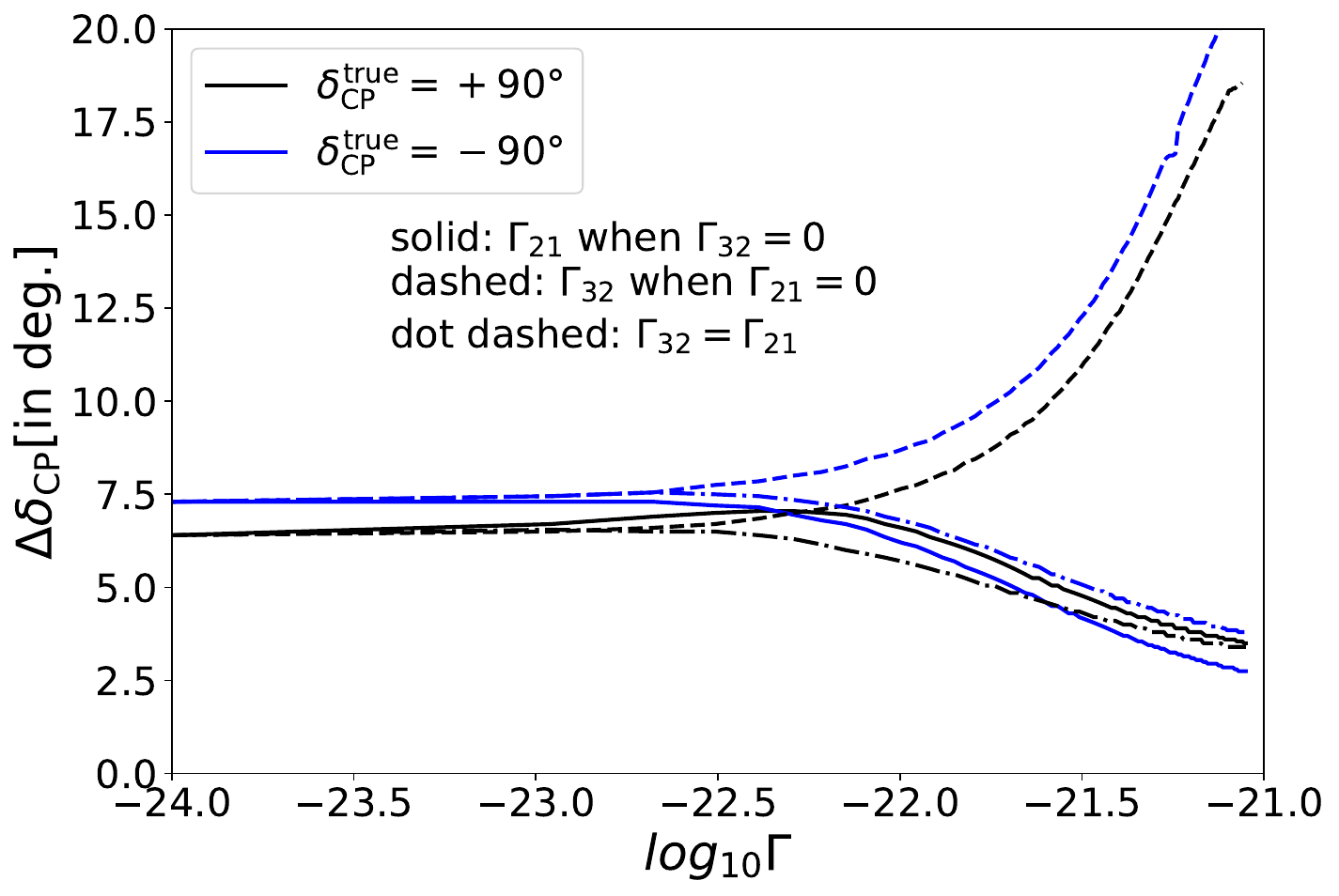}
\caption{$1\sigma$ precision of $\delta_{\rm CP}$ in \ess as a function of the decoherence parameter, for three different scenarios. Decoherence is present in both true and test data. The left (right) plot corresponds to the true values of $\delta_{\rm CP} = 0 {\rm~ and} ~180^0$ ($\pm 90^0$).   }
        \label{fig:cpv_precision}
\end{figure}

In this section, we will explore the effect of the decoherence on the uncertainty on the measurement of $\delta_{\rm CP}$ that the \ess experiment will be able to achieve. This is another crucial point since the aim of \ess will be not only to discover CPV if the next generation of LBL experiment will fail (i.e. if the $\delta_{\rm CP}$ value is not close enough to maximal value) but also to reduce the uncertainty on this parameter. It has been shown that the choice of a 360 km baseline would be optimal for measuring $\delta_{\rm CP}$ with a 1$\sigma$ uncertainty smaller than $7.5^\circ$ for all the possible values of the phase \cite{Alekou:2022emd,ESSnuSB:2023lbg}, considering that the best precision, namely $\Delta\delta_\mathrm{CP} = 5^\circ$ will be achieved for CP conserving values. This is again a result not achievable by next-generation LBL experiments. 
In Fig. \ref{fig:cpv_precision}, we show the foreseen 1$\sigma$ uncertainty on the $\delta_{\rm CP}$ measurement for the two CP conserving values ($\delta_{CP} = 0,180^\circ$, left plot) and the maximal CP violating values ($\delta_{CP} = \pm90^\circ$, right plot), by  varying the decoherence parameters up to $10^{-21}$ GeV. Even when the decoherence parameter are of the order of magnitude of the expected experimental sensitivities, the effect of \gatm and \gsol is not large enough to weaken the performances of \ess in a relevant way. In particular, on the right panel, we observe that a good $\Delta\delta_\mathrm{CP}<7.5^\circ$ can be obtained, valid for maximal $\delta_{\rm CP}$ values if $\Gamma<5\times10^{-23}$ GeV while, for CP conserving phase, $\Delta\delta_\mathrm{CP}\lesssim 7^\circ$ in the whole range of $\Gamma$'s (left panel).

Analytic considerations help in understanding the previous numerical results. Let us start from the case in which only \gsol$\ne 0$. Given the number of observed neutrino and antineutrino events $N$ and $\bar N$, their uncertainties can be written as: 
\begin{equation}
    \Delta N\sim\kappa \left| \frac{\partial P_{\mu e}}{\partial \delta_\mathrm{CP}}\right| \Delta\delta_\mathrm{CP}\,,
    \label{eq:deltaN}
\end{equation}
where $\Delta N$ can be understood as the sum of systematic and statistical uncertainties and $\kappa$ is a factor that depends on cross section and detector response. Here we have used the approximation for which, at fixed neutrino energy, the number of events is proportional to the probability and  neglected the uncertainties on the other oscillation parameters. Thus, eq. \ref{eq:deltaN} implies:
\begin{equation}
    \Delta\delta_\mathrm{CP}\propto \left|\frac{\partial P_{\mu e}}{\partial \delta_\mathrm{CP}}\right|^{-1} \, ,
\end{equation}
which, from eq. \ref{eq:probgsol}  for $\delta_{CP}=0,180^\circ$, gives:
\begin{equation}
    \Delta\delta_\mathrm{CP}\propto\frac{1}{\left|2\alpha\Delta_{31}(\Gamma_{21}L-2\sin^2\Delta_{31})-\Gamma_{21}L\sin2\Delta_{31}\right|}\,.
\end{equation}
This results clearly shows that, for small $\Gamma_{21}$, the decoherence parameter suppresses the denominator, thus increasing the uncertainty on $\delta_{\rm CP}$ while, for large $\Gamma_{21}$, $\Delta\delta_\mathrm{CP}\to0$. This behavior matches the solid curve in Fig. \ref{fig:cpv_precision}, left panel. On the other hand, for maximal CP violation, we obtain
\begin{equation}
    \Delta\delta_\mathrm{CP}\propto\frac{1}{\left|\Gamma_{21}L(\cos2\Delta_{31}+\cos2\theta_{12})+2\alpha\Delta_{31}\sin2\Delta_{31}\right|} \,.
\end{equation}
In this case, the SM contribution is already very small at the oscillation maxima; indeed, at LBL experiments, the precision around maximal phase is worse than around CP conserving values of $\delta_{\rm CP}$. Thus, the decoherence contribution simply decreases $\Delta \delta_\mathrm{CP}$, as shown in the right panel of Fig. \ref{fig:cpv_precision} (solid lines). \\
Finally, we consider the case in which only \gatm is different from zero. From eq. \ref{eq:probgatm}, for CP conserving values of the phase we obtain:
\begin{equation}
    \Delta\delta_\mathrm{CP}\propto\frac{1}{\left|\alpha\Delta_{31}\Gamma_{32}L\cos2\Delta_{31}+2\alpha\Delta_{31}\sin^2\Delta_{31}\right|}\,,
\end{equation}
from which, being $\cos2\Delta_{31}\sim-1$ around the oscillation maxima, $\Delta$\dcp~increases along with $\Gamma_{32}$. However, the effect of \gatm on the CP precision around CP conserving values is negligible because, in order to overcome the SM contribution to the CP precision, we would need a very large $\Gamma_{32}$ value, as we can observe in the left panel of Fig. \ref{fig:cpv_precision}, dashed curves. In the case of maximal CP, the uncertainty on \dcp~can be expressed as  
\begin{equation}
    \Delta\delta_\mathrm{CP}\propto\frac{1}{\left|\alpha\Delta_{31}(1-\Gamma_{32}L)\sin\Delta_{31}\right|} \,.
\end{equation}
Here the impact of \gatm on the CP precision is evident. Indeed, when the decoherence correction becomes dominant with respect to the SM probability, $\Delta\delta_{CP}\to\infty$. This behaviour is again confirmed by dashed lines of Fig. \ref{fig:cpv_precision}, right panel. However, it is important to mention that the values of the decoherence parameters for which the \dcp~precision gets relevantly influenced by the new physics are well beyond the current limits on \gsol and $\Gamma_{32}$.  As for CPV sensitivity, since the appearance probability is more influenced by \gsol than $\Gamma_{32}$, the case \gsol $ = \Gamma_{32}$ mimics the \dcp~precision curves obtained with \gsol = 0 and \gatm = 0.


\section{Summary and Conclusions}

In this work, we have performed a phenomenological study of quantum decoherence in neutrino oscillation in the context of ESSnuSB, a proposed future neutrino oscillation experiment with the primary goal of precisely measuring the Dirac CP phase $\delta_{\rm CP}$ by exploiting the features of the second oscillation maximum. We studied the capability of ESSnuSB far detector to constrain the decoherence parameters and also explored the possible effects of decoherence on the measurement of $\delta_{\rm CP}$ at ESSnuSB. Considering the neutrino as a subsystem interacting with the environment, an open quantum system framework can be utilized to probe any signature of decoherence in neutrino oscillation experiments. From eq. \ref{eq:finalprob} it appears that the effect of decoherence is to introduce damping-like terms associated with the neutrino oscillation frequencies.  Working with the simplest form of a diagonal decoherence matrix $\mathcal{D}$ and considering  {\it energy independent} decoherence parameters ($\Gamma_{ij}$, with $\Gamma_{31}=\Gamma_{21}+\Gamma_{32}-2\sqrt{\Gamma_{21}\Gamma_{32}}$), we have derived the analytical formulae for electron appearance and muon disappearance oscillation probabilities in vacuum, which is the relevant regime for the \ess experiment. Although we neglect the standard matter effect in our analytical expressions (which is a good approximation for the \ess baseline and neutrino energy), our numerical results have been obtained using exact probabilities.

Based on our analytical as well as numerical analysis, we found that the appearance channel is more sensitive towards constraining the parameter $\Gamma_{21}$, whilst sensitivity to \gatm mainly comes from the disappearance channel (whose dependence on $\Gamma_{21}$, on the other hand, is not completely negligible).
The bounds on such parameters have been obtained by means of a standard $\chi^2$ analysis performed with the help of the GloBES software. We found that,
at the $90\%$ C.L.,  \gsol $< 6.15\times 10^{-24}$ GeV and \gatm $< 1.50\times 10^{-23}$ GeV. These bounds are very much competitive with those obtained in DUNE~\cite{BalieiroGomes:2018gtd}, which amounts at $\Gamma_{21}^{\rm DUNE} < 1.2\times 10^{-23}$ GeV and $\Gamma_{32}^{\rm DUNE} < 4.7\times 10^{-24}$ GeV. Systematic errors at the level of $2\%$ will improve the previous bounds by a factor of 2-3. 

Interesting correlations appear among the atmospheric mixing angle $\theta_{23}$ and $\Gamma_{ij}$. It is well known that, in the standard oscillation scenario, \ess will have a limited capability to resolve the octant degeneracy; we observe, instead, that for $\theta_{23}$ in the lower octant, the degeneracy is resolved as soon as $\Gamma_{21}$ is larger than ${\cal O}(10^{-24})$ GeV, while even larger values of $\Gamma_{32}$ are not enough to firmly establish $\theta_{23}$ smaller than maximal mixing.
No relevant correlations have been observed between the decoherence parameters and $\delta_{CP}$.

For the latter parameter, 
we found that the capability of \ess measurement of distinguishing CP violating phase values  from $\delta_{\rm CP}=0, 180^\circ$
is robust even if decoherence exists in nature, being $\sqrt{\Delta \chi^2} \gtrsim 10$  for $\Gamma_{ij}$ in the range $[10^{-24},10^{-21}]$ GeV.
Finally, we have investigated the precision with which $\delta_{\rm CP}$ can be measured in the presence of decoherence; in the case of maximal CP violation, an uncertainty below $10^\circ$ can be maintained for  $\Gamma_{ij} \gtrsim 10^{-22}$ GeV, while becoming larger (smaller) for $\Gamma_{21}$ ($\Gamma_{32}$) above such a value.

\section*{Acknowledgements}

Funded by the European Union. Views and opinions expressed are however those of the author(s) only and do not necessarily reflect those of the European Union. Neither the European Union nor the granting authority can be held responsible for them.

We acknowledge further support provided by the following research funding agencies: Centre National de la Recherche Scientifique, France; Deutsche Forschungsgemeinschaft, Germany, Projektnummer 423761110; Ministry of Science and Education of Republic of Croatia grant No. KK.01.1.1.01.0001; the Swedish Research Council (Vetenskapsrådet) through Contract No. 2017-03934; the European Union’s Horizon 2020 research and innovation programme under the Marie Skłodowska -Curie grant agreement No 860881-HIDDeN; the European Union NextGenerationEU, through the National Recovery and Resilience Plan of the Republic of Bulgaria, project No. BG-RRP-2.004-0008-C01; as well as support provided by the universities and laboratories to which the authors of this report are affiliated, see the author list on the first page. We also acknowledge Christoph Ternes for his useful comments related to the latest bounds on decoherence parameters.

\newpage
\bibliographystyle{JHEP}
\bibliography{decoherence_ref.bib, references,ref_deco.bib}

\end{document}